\documentclass[twocolumn,showpacs,amsmath,amssymb,bibliography]{revtex4}

\usepackage{graphicx,amsmath,amssymb,bm,multirow}
\usepackage{amsthm}
\usepackage{amsfonts}
\usepackage{dcolumn}
\usepackage{bm}
\usepackage{mathrsfs}
\usepackage[usenames,dvipsnames]{color}
\usepackage{amsfonts,mathrsfs,amssymb}
\usepackage{xspace}

\usepackage[colorlinks,linkcolor=blue,anchorcolor=blue,citecolor=blue]{hyperref}

\newcommand{\bq}{{\mathbf{q}}}

\newcommand{\beq}{\begin{equation}}
\newcommand{\eeq}{\end{equation}}
\newcommand{\beqn}{\begin{eqnarray}}
\newcommand{\eeqn}{\end{eqnarray}}
\newcommand{\bsub}{\begin{subequations}}
\newcommand{\esub}{\end{subequations}}
\newcommand{\bpm}{\begin{pmatrix}}
\newcommand{\epm}{\end{pmatrix}}

\newcommand{\te}{\text{e}}
\newcommand{\ti}{\text{i}}

\newcommand{\red}[1]{\textcolor{red}{#1}}
\newcommand{\green}[1]{\textcolor{green}{#1}}
\newcommand{\blue}[1]{\textcolor{blue}{#1}}

\begin{document}

\title{Octupole correlations in low-lying states of $^{150}$Nd and $^{150}$Sm
and their impact on neutrinoless double-beta decay}
 \author{J. M. Yao\footnote{On leave from School of Physical Science and Technology, Southwest University, 400715 Chongqing, China}}
\affiliation{Department of Physics and Astronomy, University of North Carolina, Chapel Hill, NC 27516-3255, USA}
\author{J. Engel}
\affiliation{Department of Physics and Astronomy, University of North Carolina,
Chapel Hill, NC 27516-3255, USA}


\begin{abstract}
We present a generator-coordinate calculation, based on a relativistic
energy-density functional, of the low-lying spectra in the isotopes $^{150}$Nd
and $^{150}$Sm and of the nuclear matrix element that governs the neutrinoless
double-beta decay of the first isotope to the second.  We carefully examine the
impact of octupole correlations on both nuclear structure and the double-beta
decay matrix element.  Octupole correlations turn out to reduce quadrupole
collectivity in both nuclei.  Shape fluctuations, however, dilute the effects
of octupole deformation on the double-beta decay matrix element, so that the
overall octupole-induced quenching is only about 7\%.

\end{abstract}

\maketitle


\section{Introduction}

The nucleus $^{150}$Nd has been the active isotope in double-beta
($\beta\beta$) decay experiments \cite{Argyriades09}, and may be again in the
future \cite{Bongrand2015}.  It has $N=90$ neutrons and is part of a set of
isotones in which nuclear structure changes rapidly as neutrons are added or
removed.  Ref.\ \cite{Casten2001} showed that the low-lying states of $^{150}$Nd
and other $N=90$ isotones are close to the predictions of the X(5) model, which
describes the critical point of a first order phase transition from spherical
harmonic vibrator to axially-deformed rotor. A significant amount of work on
nuclear phase transitions in the $N\simeq90$ isotones followed this
discovery \cite{Krucken2002,Meng05,Sheng05,Niksic07,Guzman07,Rodriguez08,Robledo08,Li09}.
We concern ourselves here, however, with a different aspect of structure in
$^{150}$Nd: octupole correlations, suggested by the observation of low-lying
negative-parity states and fast $E3$
transitions \cite{Urban87,Friedrichs92,Bvumbi13}.
Refs.\ \cite{Nazarewicz92,Garrote98,Babilon05,Minkov06,Zhang10,Robledo11,Guzman2012}
applied a wide variety of models with octupole shape degrees of freedom to
nuclei with $N\simeq90$.  The models indicated that the proton
1h$_{11/2}$--2d$_{5/2}$ and neutron 1i$_{13/2}$--2f$_{7/2}$
orbitals near the Fermi levels are responsible for the strong octupole
correlations.  And very recently the authors of
Refs.\ \cite{Nomura2014,Nomura2015} used the $sdf$ interacting Boson model
(IBM), with Hamiltonian parameters determined from self-consistent mean-field
calculations, to successfully describe the low-lying states of $N\simeq90$
nuclei.  These studies imply that the low-lying states of $^{150}$Nd are
dominated by the quadrupole-octupole collective excitations and that the EDF
approaches provide good basis states for them.

The $sdf$ IBM calculation\ \cite{Nomura2014,Nomura2015} can be regarded ---
very roughly speaking --- as something like an EDF-based shell-model
calculation with the model space truncated to states constructed from nucleon
pairs with angular momentum $J = 0, 2$, and 3.  It obviously contains
correlations beyond those of mean-field theory. In this paper, we carry out a
symmetry-projected beyond-mean-field calculation, using the Generator Coordinate
Method (GCM) to study the low-lying states of $^{150}$Nd and $^{150}$Sm and to
quantify the effects of octupole correlations on the matrix element that
governs neutrinoless double-beta ($0\nu\beta\beta$) decay from the ground state
of the first to that of the second.  Our starting point is a self-consistent
relativistic mean-field (RMF) calculation
\cite{Reinhard89,Ring96,Vretenar05,Meng06} with constraints on both quadrupole
and octupole mass moments, part of a framework that has already been used to
study static octupole deformation in nuclear ground states
\cite{Geng07,Guo10,Zhang10,Lu12,Li13}. The GCM approach we take here (also
known as multi-reference covariant density-functional theory) allows us to go
beyond mean-field theory, however, by including dynamical correlations
associated with symmetry restoration and shape
fluctuations \cite{Niksic06,Yao13,Wu14,Wu15,Yao10,Yao11,Yao14}.  Octupole shape
fluctuations \cite{Yao15-Ra,Zhou16} have not yet been studied extensively.
Previous work on the $\beta\beta$ decay of $^{150}$Nd has shown that the matrix
elements are sensitive to quadrupole
deformation \cite{Hirsch95,Chaturvedi2008,Fang2010,Rodriguez2010,Mustonen2013,Song2014,
Terasaki2014,Yao2015}.  It is the effect of octupole correlations that we
address here.

This paper is organized as follows: Section \ref{Formalism} briefly presents
the RMF theory that we use to generate reference states, the GCM and several
projection techniques that we apply to nuclear collective quadrupole and
octupole excitations, and the formulae for computing the matrix elements of the
operator responsible for $0\nu\beta\beta$ decay.  Section \ref{result} presents
our results for the structure of low-lying states in $^{150}$Nd and $^{150}$Sm,
and for the $0\nu\beta\beta$ matrix elements, which we compare with those of
previous studies that neglect octupole degrees of freedom.  Section
\ref{summary} summarizes our findings.

\section{The model}
\label{Formalism}

 \subsection{Generating mean-field reference states in collective coordinate space}
The first step in the GCM procedure is to generate a set of collective reference
(or basis) states $\vert q\rangle$. We do so by carrying out constrained
mean-field calculations based with a relativistic point-coupling energy-density
functional (EDF) \cite{Nikolaus92,Burvenich02,Zhao10}:
\begin{equation}
\label{EDF}
{{E}}_{\textnormal{RMF}} =
 \int d{\bm r }~{\mathcal{E}_{\textnormal{RMF}}}[\rho, j] \,,
\end{equation}
where $\mathcal{E}_{\rm RMF}$ is defined as
\begin{align}
{\mathcal{E}_{\textnormal{RMF}}}[\rho, j]
&= \sum_{k,\tau}v_{k,\tau}^2 \,{\bar{\psi}_{k,\tau} (\bm{r}) \left( -i\bm{\gamma}
\bm{\nabla} + m\right )\psi_{k,\tau}(\bm{r})}  \nonumber \\
 &+   \frac{\alpha_0}{2}\rho^2+\frac{\beta_0}{3}\rho^3 +
  \frac{\gamma_0}{4}\rho^4+\frac{\delta_0}{2}\rho\triangle \rho
  \nonumber \\
 &
 + \frac{\alpha_1}{2}j_\mu j^\mu  + \frac{\gamma_1}{4}(j_\mu j^\mu)^2
 +  \frac{\alpha_{11}}{2}\tilde{j}^{\mu} \tilde{j}_\mu \nonumber \\
 &+
       \frac{\delta_1}{2}j_\mu\triangle j^\mu
       +\frac{\delta_{11}}{2}
       \tilde{j}^\mu \triangle \tilde{j}_{\mu}  \nonumber \\
       & + \frac{\alpha_{10}}{2}
       \tilde{\rho}^2+\frac{\delta_{10}}{2}\tilde{\rho}\triangle
       \tilde{\rho} +\frac{1}{4}e(j_\mu-\tilde{j}_\mu) A^\mu \,.
\label{eq:EDF}
\end{align}
Here $\tau=1$ (neutron) or $-1$ (proton), $\psi_{k,\tau}$ is the Dirac wave
function for the $k^{\rm th}$ nucleon with isospin $\tau$, and $A^\mu$ is the
electromagnetic field.  The functional contains eleven constants $\alpha$,
$\beta$, $\gamma$ and $\delta$.  The local isoscalar and isovector densities
$\rho$ and $\tilde{\rho}$, and the corresponding isoscalar and isovector
currents $j_\mu$ and $\tilde{j}_\mu$ are
\bsub\begin{align}
\label{dens_1}
\rho({\bm r}) &=\sum_{k,\tau} v_{k,\tau}^2 \, \bar{\psi}_{k,\tau}({\bm r})
             \psi _{k,\tau}({\bm r})\,,  \\
\label{dens_2}
\tilde{\rho}({\bm r}) &=\sum_{k,\tau} \tau \, v_{k,\tau}^2 ~
\bar{\psi}_{k,\tau}({\bm r})\psi _{k,\tau'}({\bm r}) \,,  \\
\label{dens_3}
j^{\mu}({\bm r}) &=\sum_{k,\tau} v_{k,\tau}^2 ~\bar{\psi}_{k,\tau}({\bm r})
        \gamma^\mu\psi _{k,\tau}({\bm r}) \,,  \\
\label{dens_4}
\tilde{j^{\mu}}({\bm r}) &=\sum_{k,\tau} \tau\, v_{k,\tau}^2 \,
\bar{\psi}_{k,\tau}({\bm r})
     \gamma^\mu \psi _{k,\tau}({\bm r}) \,.
\end{align}
\esub These quantities are calculated in the {\it no-sea} approximation, i.e.\
with the summation in Eqs.\ \eqref{dens_1} -- \eqref{dens_4} running over all
states with $v^2_{k,\tau}>0$, where the $v_{k,\tau}^2$ is the occupation
probability for the $k^{\rm th}$ nucleon of type $\tau$ in the BCS wave function
\beq
\label{eq:bcs}
\vert q\rangle = \prod_{k>0,\tau}(u_{k,\tau}+ v_{k,\tau} c^\dagger_{k,\tau}
c^\dagger_{\bar k,\tau})\vert
0\rangle \,.
\eeq
The numbers $u_{k,\tau}$ and $v_{k,\tau}$ obey $u^2_{k,\tau}+v^2_{k,\tau}=1$.
The operator $c^\dagger_{k,\tau}$ creates a nucleon of type $\tau$ in the
single-nucleon state $k$.  The corresponding spinor $\psi_{k,\tau}$ is
determined by the variational principle
 \begin{equation}\label{Dirac}
 \delta \langle q\vert \hat H - \sum_{\tau} \lambda_\tau \hat N_\tau -
 \sum_{\lambda=1, 2,3} C_\lambda (\hat Q_{\lambda0} - q_\lambda)^2  \vert
 q\rangle=0 \,,
\end{equation}
with the Lagrange multipliers $\lambda_\tau$ fixed by the constraints $\langle q
\vert \hat N_1 \vert q\rangle=N$ and $\langle q \vert \hat N_{-1} \vert
q\rangle=Z$.
The deformation parameters $\beta_\lambda$ ($\lambda=2, 3$) are related to the
mass multipole moments by
\begin{equation}\label{deformation}
  \beta_\lambda \equiv \dfrac{4\pi}{3A R^\lambda} q_\lambda, \quad R=1.2A^{1/3}
  \,,
\end{equation}
with $A$ representing the mass number of the nucleus under consideration.

We iteratively solve the Dirac equation derived from (\ref{Dirac}) by expanding
the Dirac spinors $\psi_{k,\tau}$ in a basis of single-particle oscillator
states within 12 shells \cite{Gambhir90}. As Eq.\ \eqref{eq:bcs} indicates, we
treat pairing correlations in the BCS approximation, with a density-independent
zero-range pairing interaction \cite{Krieger90}. We always employ the
relativistic energy density functional PC-PK1 \cite{Zhao10}.  Previous
symmetry-projected GCM studies \cite{Niksic07,Song2014} have shown that the
low-lying states produced by the PC-PK1 and the PC-F1 \cite{Burvenich02} are
close to each other in energy, suggesting that reasonable changes in the
particle-hole structure of the energy-density functional will not produce major
changes in low-lying structure.  The pairing functional, however, has been
shown to have a significant effect in $^{150}$Nd, on both in its collective
structure \cite{Li11} and its matrix element for neutrinoless double beta decay
\cite{Song2014}.  Here, as in Ref.\ \cite{Song2014} we choose to fit the
pairing strengths to the average pairing gaps produced by a separable
finite-range pairing force at the mean-field energy minimum \cite{Tian2009}.
The procedure leads to pairing gaps that are similar to those obtained both
from the Gogny functional and experiment.

 \subsection{Symmetry restoration and configuration mixing}

We construct physical state vectors $\vert J^\pi_\alpha\rangle$ by superposing
projected mean-field reference states:
\begin{equation}\label{gcmwf}
\vert J^\pi_\alpha\rangle
=\sum_{q} f^{J\pi \alpha}_q   \vert JM\pi NZ; q\rangle\,,
\end{equation}
where $\vert JM\pi NZ; q\rangle\equiv\hat P^J_{M0} \hat P^N\hat P^Z \hat
P^\pi\vert q\rangle$, with the ``0'' in the first projector corresponding to
the intrinsic quantum number $K$ (which will be zero for all our states) and
the collective coordinate $q$ standing for the intrinsic deformation parameters
$(\beta_2, \beta_3)$ of the reference states.  The $\hat P$'s are projection
operators onto states with well-defined angular momentum $J$ and its
$z$-component $M$, parity ($\pi=\pm$), and neutron and proton number ($N, Z$)
\cite{Ring1980}. [$N$, $Z$, and $K$ are implied on the left-hand side of Eq.\
\eqref{gcmwf}]. The weight functions $f^{J\pi\alpha}_q$, where $\alpha$ is a
simple enumeration index, are solutions to the Hill-Wheeler-Griffin
equation \cite{Hill53,Griffin57}
\begin{equation}\label{HWG}
\sum_{q'} \left[  \mathscr{H}^{J\pi}_{q, q'}-E^{J\pi}_{\alpha}
\mathscr{N}^{J\pi}_{q, q'}\right]f^{J\pi \alpha}_{q'}=0\,,
\end{equation}
where the Hamiltonian kernel $ \mathscr{H}^{J\pi}_{q, q'}$ and the norm kernel
$\mathscr{N}^{J\pi}_{q, q'}$ are given by
\begin{equation}
\label{kernel}
\left\{ \begin{array}{c} \mathscr{H}\\ \mathscr{N}
\end{array} \right\}^{J\pi}_{q,q'}
=\langle q \vert  \left\{ \begin{array}{c} \hat{H}\\1 \end{array} \right\}
\hat P^J_{00} \hat P^N\hat P^Z  \hat P^\pi\vert q'\rangle \,,
\end{equation}
To solve Eq.\ (\ref{HWG}), we first diagonalize the norm kernel $\mathscr{N}$
and then use the non-zero eigenvalues and corresponding eigenvectors to
construct the ``natural basis'' \cite{Ring1980,Guzman2002,Yao10}.  We re-diagonalize the
Hamiltonian in that basis to obtain the states $\vert J^\pi_\alpha \rangle$ and
the energies $E^{J\pi}_{\alpha}$.

Because we begin with an energy functional rather than a Hamiltonian, we need a
prescription for the off-diagonal matrix elements of $\mathscr{H}$.  Following
standard practice, we simply replace the diagonal density in the functional by
the transition density.  Though the prescription brings with it spurious
divergences and ``steps'' \cite{Lacroix09,Duguet09}, it does not produce an
unresolvable ambiguity when used together with the relativistic EDF in Eq.\
(\ref{EDF}), which contains only integer powers of the density.  We exclude
exchange terms but avoid numerical instabilities in particle-number projection
at the gauge angle $\phi=\pi/2$ by setting $L$ to 9 in Fomenko's expansion
\cite{Fomenko1970}.  Refs.\
\cite{Guzman2002,Niksic06,Bender08,Rodriguez10,Yao10} contain detailed
discussions of beyond-mean-field calculations with energy-density functionals.

 \subsection{Nuclear matrix element for $0\nu\beta\beta$ decay}

The $0\nu\beta\beta$ decay nuclear matrix element is
\beqn
\label{NME:formula}
M^{0\nu} \nonumber &=&\dfrac{4\pi R}{g^2_A(0)}\int\int d^3 x_1d^3 x_2
\int\frac{d^3 q}{(2\pi)^3}\frac{\te ^{\ti \bm q\cdot(\bm x_1-\bm
x_2)}}{q}\notag\nonumber\\
&\times& \sum_m\frac{\langle 0^+_F\vert \mathcal {J}_{\mu}^\dagger(\bm
x_1)|m\rangle\langle m|\mathcal {J}^{\mu\dagger}(\bm x_2)\vert
0^+_I\rangle}{q+E_m-(E_I+E_F)/2},
\eeqn
where ${\cal J}^\dagger_{\mu}$ is the charge-changing nuclear current operator
\cite{Avignone08} and $q$ is the momentum transferred from leptons to
nucleons.  The nuclear radius $R=1.2A^{1/3}$ makes the matrix element
dimensionless.  In the closure approximation and with the GCM state vectors
from Eq.\ (\ref{gcmwf}) as ground states $\vert 0^+_{I/F}\rangle$ of the
initial and final nuclei, we obtain
\begin{equation}
M^{0\nu} =  \sum_{q_I,q_F}f^{0^+_I}_{q_I} f^{0^+_F}_{q_F}
\langle q_F\vert \hat {\cal O}^{0\nu}  \hat P^{J=0}_{00} \hat P^N\hat P^Z  \hat
P^{\pi=+}\vert q_I\rangle \,,
\end{equation}
with the transition operator given by
\beqn
\hat {\cal O}^{0\nu}
&=&\dfrac{4\pi R}{g^2_A(0)}\int\frac{d^3 q}{(2\pi)^3}\int\int d^3 x_1d^3
x_2\frac{\te ^{\ti \bm q\cdot(\bm x_1-\bm x_2)}}{q (q+E_d)} \nonumber\\
&\times& [\mathcal {J}_{\mu}^\dagger(\bm x_1) \mathcal {J}^{\mu\dagger}(\bm
x_2)] \,,
\eeqn
and $E_d$ set to $1.12 A^{1/2} \simeq 13.72$ Mev \cite{Haxton84}.

The operator $[\mathcal {J}_{\mu}^\dagger(\bm x_1) \mathcal
{J}^{\mu\dagger}(\bm x_2)]$, when Fourier transformed, contains the terms
\cite{Song2014},
\begin{eqnarray}
\label{twocurrentR}
VV: &&g_V^2(\bm q^2)\left(\bar\psi\gamma_\mu\tau_-\psi\right)^{(1)}\left(\bar\psi\gamma^\mu\tau_-\psi\right)^{(2)} \\
AA: &&g_A^2(\bm
q^2)\left(\bar\psi\gamma_\mu\gamma_5\tau_-\psi\right)^{(1)}\left(\bar\psi\gamma^\mu\gamma_5\tau_-\psi\right)^{(2)}\nonumber \\
AP: &&2g_A(\bm q^2)g_P(\bm q^2)\left(\bar\psi\bm
\gamma\gamma_5\tau_-\psi\right)^{(1)}\left(\bar\psi \bm
q\gamma_5\tau_-\psi\right)^{(2)}\nonumber\\
PP: &&g_P^2(\bm q^2)\left(\bar\psi \bm
q\gamma_5\tau_-\psi\right)^{(1)}\left(\bar\psi \bm
q\gamma_5\tau_-\psi\right)^{(2)}\nonumber\\
MM: &&g_M^2(\bm q^2)\left(\bar\psi\frac{\sigma_{\mu
i}}{2m_N}q^i\tau_-\psi\right)^{(1)}\left(\bar\psi\frac{\sigma^{\mu
j}}{2m_N}q_j\tau_-\psi\right)^{(2)}\,,\nonumber
\end{eqnarray}
where $\tau_-$ is the isospin lowering operator that changes neutrons into
protons, $\sigma_{\mu\nu}=\frac{\ti}{2}\left[\gamma_\mu,\gamma_\nu\right]$, and
$V,A,P,M$ denote the vector, axial vector, pseudoscalar, and magnetic pieces of
the one-nucleon current.  Following Ref.\ \cite{Simkovic1999}, we take the form
factors $g_V(\bm q^2)$, $g_A(\bm q^2)$, $g_M(\bm q^2),$ and $g_P(\bm q^2)$ to
be
\bsub\beqn
g_V(\bq^2) &=&\dfrac{g_V(0)}{(1+\bq^2/\Lambda^2_V)^2},\\
g_A(\bq^2) &=&\dfrac{g_A(0)}{(1+\bq^2/\Lambda^2_A)^2}, \\
g_P(\bq^2) &=& g_A(\bq^2)\dfrac{2m_N }{\bq^2+m^2_\pi} (1-\dfrac{m^2_\pi}{\Lambda^2_A}),\\
g_M(\bq^2) &=& (\mu_p-\mu_n) g_V(\bq^2),
\eeqn\esub
with $g_V(0) = 1.0$, $g_A(0) = 1.254$, $\mu_p-\mu_n=3.70$, $\Lambda^2_V = 0.710$ (GeV)$^2$, $\Lambda_A = 1.09$ GeV,
$m_N=0.93827$ GeV and $m_\pi=0.13957$ GeV.  For the sake of simplicity, we
neglect short-range correlations.

We include, alongside the generator coordinates from Ref.\ \cite{Song2014}, the
octupole deformation parameter $\beta_{3}$.  The parity breaking (and
subsequent projection) and the larger number of reference states caused by the
inclusion of octupole deformation increase computing time but otherwise cause
no problems in our calculation.  We initially include 50 reference states
with $\beta_3>0$.  From this set, 29 natural states turn out to sufficient to
include essentially all the contributions of the original 50 states to both
structure properties and $0\nu\beta\beta$ decay matrix elements.

\section{Results and discussion}
\label{result}

\begin{figure}[]
\centering
\includegraphics[width=\columnwidth]{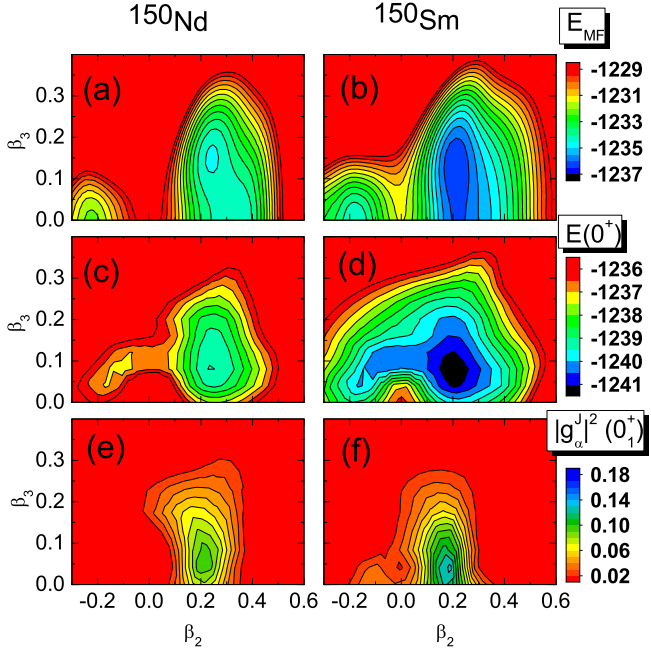}
\vspace{-0.5cm}
\caption{(Color online) Mean-field energy surfaces for $^{150}$Nd (a) and
$^{150}$Sm (b), projected energy surfaces for $^{150}$Nd (c) and $^{150}$Sm (d),
and the square of the collective ground-state wave function for $^{150}$Nd (e)
and $^{150}$Sm (f), all in the $\beta_2$-$\beta_3$ plane.  }
\label{PES_WF}
\end{figure}

Figure \ref{PES_WF} shows the mean-field and quantum-number-projected energy
surfaces, as well as the collective wave functions $\vert g^J_\alpha(q)\vert^2$,
for the ground states of $^{150}$Nd and $^{150}$Sm. The collective wave
functions, defined as $g^{J\pi}_\alpha (q) \equiv \sum_{q'}
\left[\mathscr{N}^{J\pi}_{q, q'}\right]^{1/2}f^{J\pi\alpha}_{q'}$, provide
information about the importance of deformation in the state $\vert
J^\pi_\alpha\rangle$.  The mean-field energy surfaces in both nuclei around the
quadrupole-deformed minima with $\beta_2$ around 0.2 are almost flat in the
octupole direction.  This kind of surface often signifies a critical point
symmetry \cite{Meng05,Niksic07,Li09}.  Our surface, however, is flat only before
projection of the states that determine it onto the subspace with $J^\pi=0^+$
and well-defined $N$ and $Z$; after projection it shows pronounced minima around
$\beta_3\sim0.1$.  This is a genuine effect, arising from the restoration of the
symmetries spontaneously broken at the mean-field level.  In addition, valleys
connect the prolate and oblate minima through octupole shapes in both nuclei.
As a result, the collective wave functions are shifted towards smaller
quadrupole deformation via the octupole coordinate; quadrupole collectivity is
thus reduced and octupole shape fluctuations are large.

\begin{figure}[tb]
\centering
\includegraphics[width=\columnwidth]{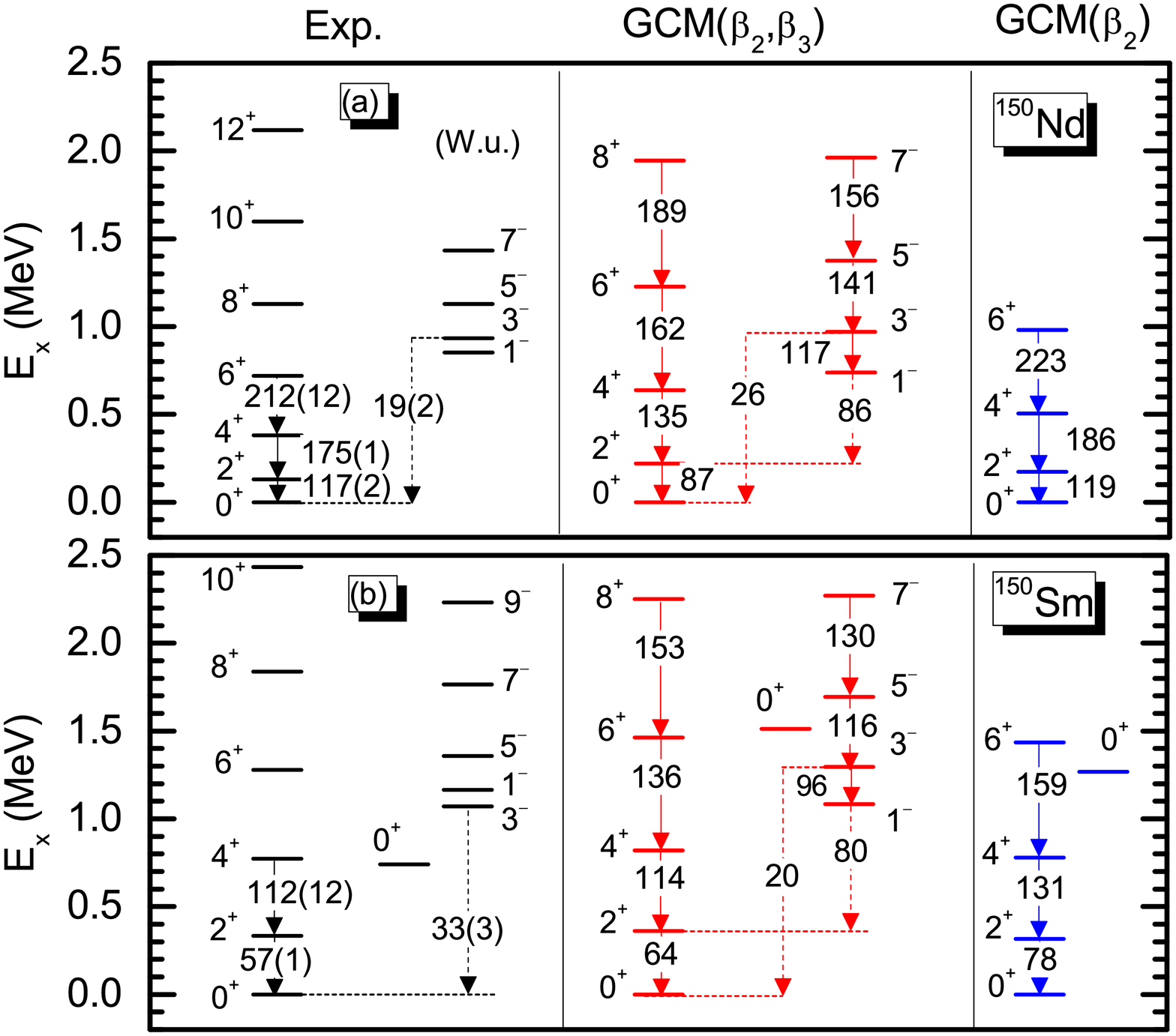}
\caption{(Color online) Low-energy spectra of $^{150}$Nd and $^{150}$Sm.  The
numbers on arrows are $E2$ (solid line) and $E3$ (dashed line) transition
strengths, in Weisskopf units.  Data are from Ref.\ \cite{NNDC}.}
\label{spectra}
\end{figure}

\begin{figure}[tb]
\centering
\includegraphics[width=\columnwidth]{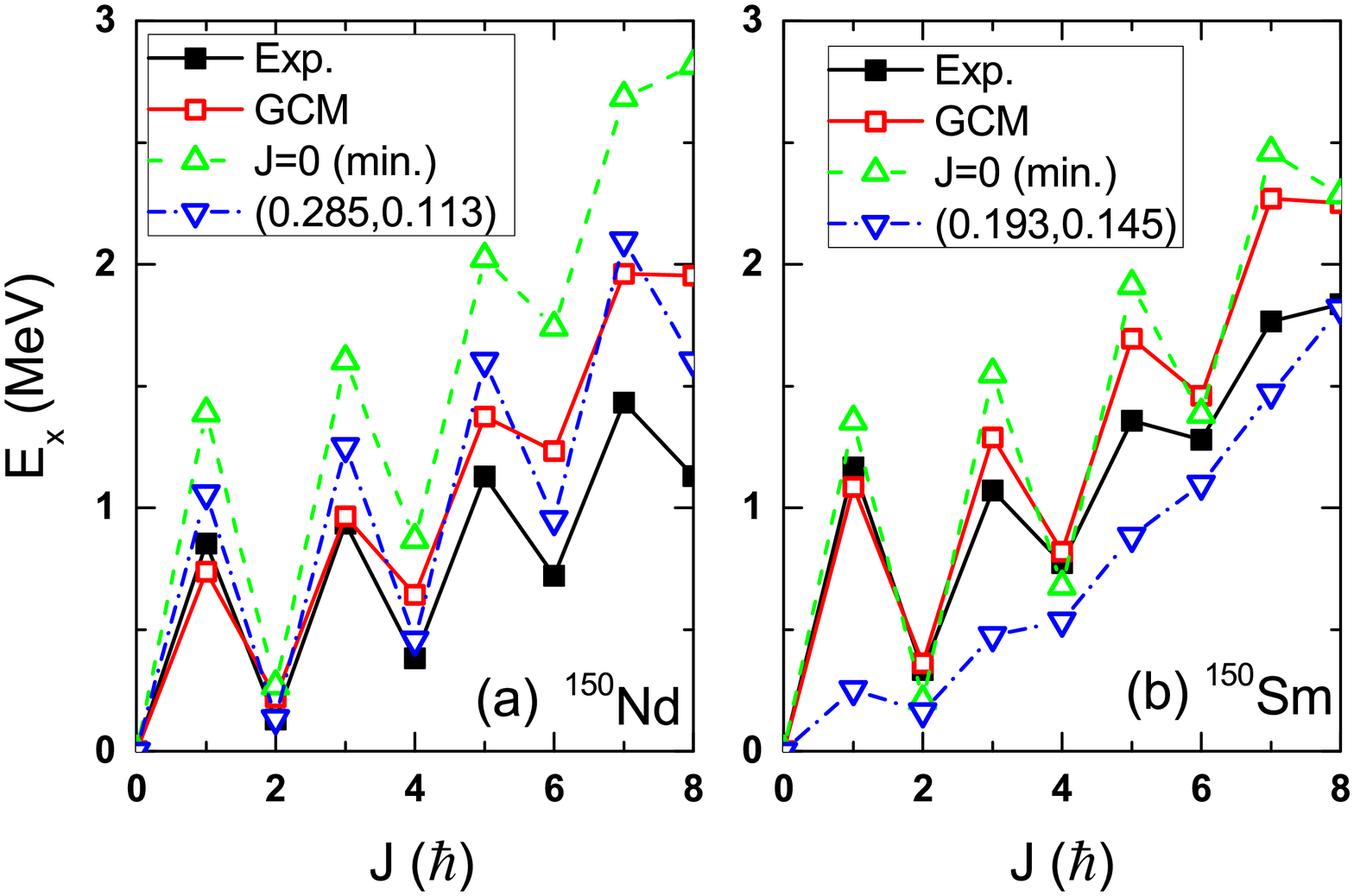}
\caption{(Color online) Excitation energies of parity doublet states in
$^{150}$Nd (a) and $^{150}$Sm (b). The available data ($\blacksquare$) are
compared with the GCM results ($\red{\square}$) and the results produced by the
single-configuration of $J=0$ energy minimum ($\green{\vartriangle}$) and by
the configuration with deformation parameters determined by measured
transition strengths $B(E2: 0^+_1 \to 2^+_1)$ and $B(E3: 0^+_1 \to 3^-_1)$
($\blue{\triangledown}$) \cite{NNDC}.}
\label{staggering}
\end{figure}

Figure \ref{spectra} shows the low-lying energy spectra in $^{150}$Nd and
$^{150}$Sm. The octupole degree of freedom reduces the $E2$ transition
strengths between positive-parity states significantly in both nuclei.  It
worsens the agreement in $^{150}$Nd but improves it in $^{150}$Sm.  Our GCM
describes the negative-parity band built on the $1^-$ state rather well,
despite overestimating the transition strength $B(E3: 0^+_1 \to 3^-_1)$ in
$^{150}$Nd and underestimating it in $^{150}$Sm.

Figure \ref{staggering} compares the GCM excitation energies with those
calculations that use a single symmetry-projected BCS state, either the one that
corresponds to the $J=0$ energy minimum or the one with deformation parameters
determined by the experimental $B(E2: 0^+_1 \to 2^+_1)$ and $B(E3: 0^+_1 \to
3^-_1)$ values \cite{NNDC}.  The GCM results that include the
configuration-mixing effect are in much better agreement with the data than
those based on a single BCS state.  As spin increases, however, the GCM
increasingly over-predicts the data, indicating that some important correlations
are missing.  Time-reversal-symmetry-breaking reference states, produced in a
cranking calculation, would likely lower the energies of high-spin states
\cite{Borrajo15}.

\begin{figure}[tb]
\centering
\includegraphics[width=\columnwidth]{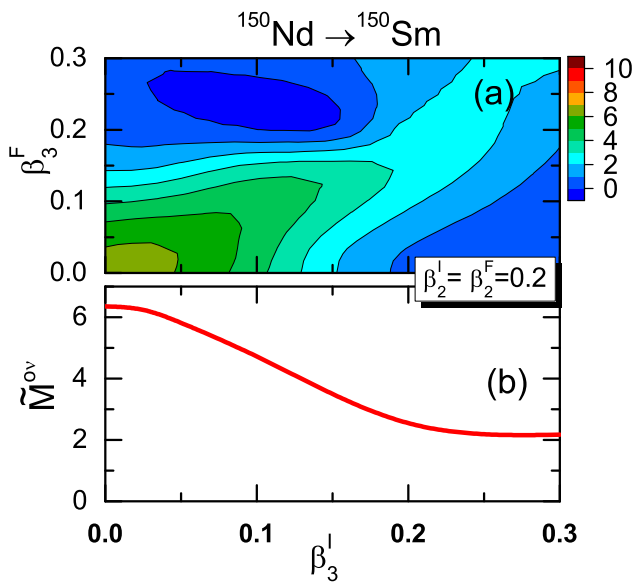}
\caption{(Color online) Normalized nuclear matrix elements $\tilde
M^{0\nu}$($q_I, q_F$) for the neutrinoless double-beta decay of $^{150}$Nd,
where $\{q\}\equiv\{\beta_2,\beta_3\}$. Panel (a) plots $\tilde M^{0\nu}$
versus the initial and final octupole deformation parameters, with the
quadruple deformation parameters $\beta^I_2$ and $\beta^F_2$ fixed at 0.2 Panel
(b) plots the same quantity with the restriction $\beta^I_3=\beta^F_3$.}
\label{NME:deformation}
\end{figure}

\begin{figure}[]
\centering
\includegraphics[width=\columnwidth]{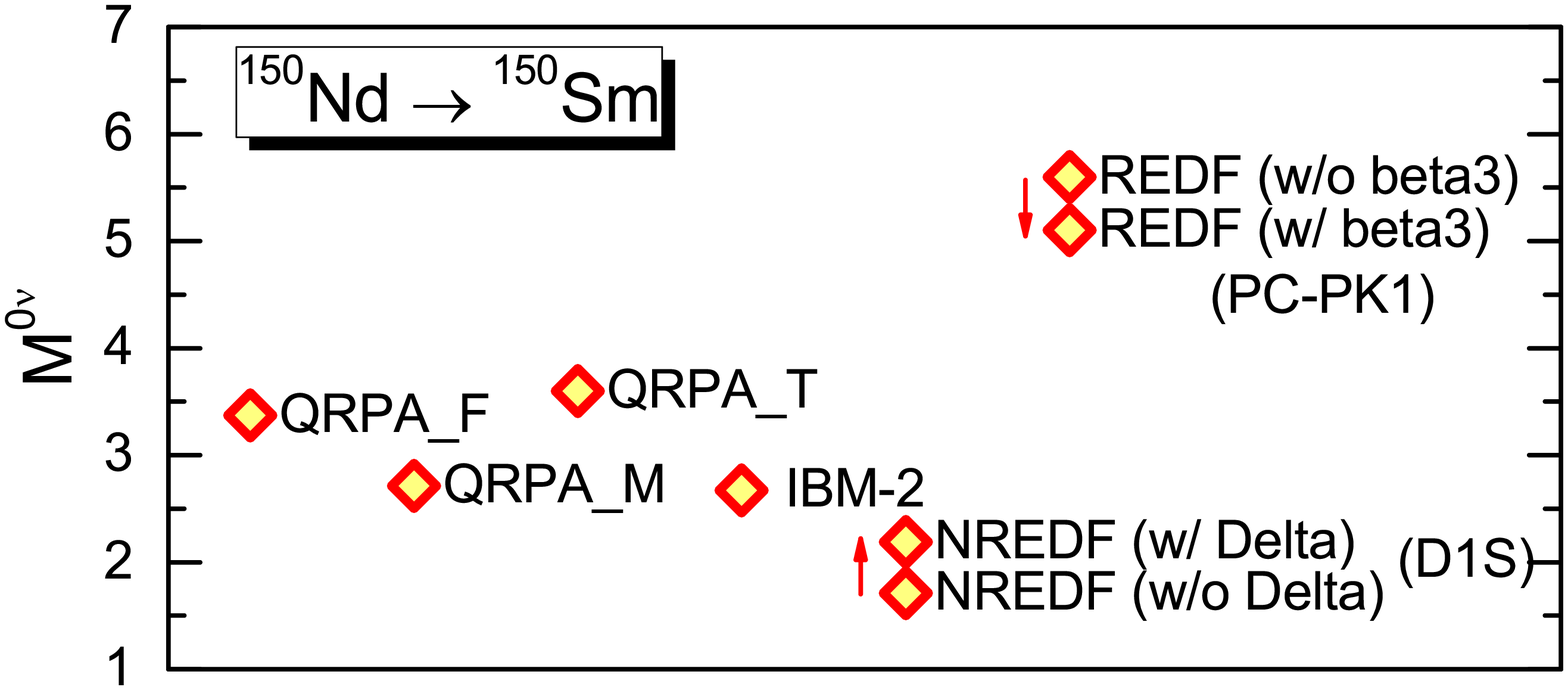}
\caption{(Color online) The final matrix element $M^{0\nu}$ from the GCM
calculation with and without \cite{Song2014} octupole shape fluctuations (REDF)
and those of the QRPA (``QRPA\_F" \cite{Fang15}, ``QRPA\_M"
\cite{Mustonen2013}, ``QRPA\_T" \cite{Terasaki2014}), the IMB-2
\cite{Barea2009}, and the non-relativistic GCM, based on the Gogny D1S  interaction,
with \cite{Vaquero2013} and without \cite{Rodriguez2010} pairing fluctuations.}
\label{NME:GCM}
\end{figure}

Figure \ref{NME:deformation} displays the normalized $0\nu\beta\beta$ matrix
element between reference states, which we denote by
$\tilde{M}^{0\nu}(q_I,q_F)$:
 \beqn
\label{NME:norm}
&&\tilde{M}^{0\nu}(q_I, q_F)\nonumber\\
&\equiv& \dfrac{\langle q_F|\hat{\mathcal O}^{0\nu}\hat P^{J=0} \hat
P^{N_I}\hat P^{Z_I}P^{\pi=+} | q_I\rangle}
{\sqrt{\mathscr{N}^{0+}_{q_I,q_I} \mathscr{N}^{0+}_{q_F,q_F} } } \,,
\eeqn
with the norms $\mathscr{N}$ for each nucleus defined in Eq.\ \eqref{kernel}.
The function $\tilde{M}^{0\nu}(q_I, q_F)$ represents the contribution of
particular initial and final configurations to the full matrix element.  Panel
(a) of Fig.\ \ref{NME:deformation} plots the function in the $\beta^I_3,
\beta^F_3$ plane, with $\beta^I_2$ and $\beta^F_2$ fixed at 0.2, the value that
minimizes the energy in both nuclei.  The figure shows that unequal octupole
deformation in the two nuclei causes a rapid drop in the $0\nu\beta\beta$
matrix element.  Panel (b) of Fig.\ \ref{NME:deformation} extracts the behavior
of $\tilde{M}^{0\nu}$ from the diagonal of panel (a), where the octupole
deformation is the same size in both nuclei.  Increasing deformation causes
even this diagonal contribution to drop, from 6.4 to 2.2 as $\beta_3$ increases
to 0.3 At the configurations that minimize the projected energies, with both
values of $\beta_2$ about 0.2 and both values of $\beta_3$ about 0.1,
$\tilde{M}^{0\nu}$ is 4.76. At the configuration that best fits the
experimental $B(E2: 0^+_1 \to 2^+_1)$ and $B(E3: 0^+_1 \to 3^-_1)$ values,
corresponding to deformation parameters $\beta_2^I= 0.285, \beta_3^I= 0.113,
\beta^F_2=0.193, \beta_3^F=0.145$, $\tilde{M}^{0\nu}$ is only 1.38.

As already discussed in Refs.\ \cite{Song2014,Yao2015}, $\tilde M^{0\nu}$ near
spherical shapes is much larger than predicted by the Gogny D1S interaction
\cite{Rodriguez2010}.  The difference arises at least in part from a difference
in average pairing gaps, which for the neutrons in $^{150}$Nd and $^{150}$Sm
are about 30\% larger here than in Ref.\ \cite{Rodriguez2010} (even though the
gaps are similar at the mean-field minima).

When all configurations are appropriately combined, we obtain a final value for
the matrix element $M^{0\nu}(0^+_1 \to 0^+_1)$ of 5.2, which is just $7\%$
smaller than the result 5.6 obtained without octupole deformation
\cite{Song2014}. (The contributions from the $VV, AA, AP, PP$, and $MM$ terms
are 1.03, 4.87, $-1.65$, 0.70, and 0.21, respectively).  The small reduction,
significantly less than what would result from the use of the single
configuration in each nucleus that minimizes the energy (4.76) shows that shape
fluctuations wash out the effects of octupole deformation.  For the
$0\nu\beta\beta$ decay to the excited $0^+$ state in $^{150}$Sm, we find
$M^{0\nu}(0^+_1 \to 0^+_2)=0.72$.

Figure \ref{NME:GCM} compares the ground-state to ground-state matrix elements
$M^{0\nu}(0^+_1 \to 0^+_1)$ from several models. Our relativistic EDF-based GCM
result is still about twice those of the deformed quasiparticle random phase
approximation (QRPA) and the interacting boson model (IBM), and about three
times that of the non-relativistic Gogny-based GCM.  A more careful study of
shell structure and pairing will help resolve the last discrepancy.  And we can
expect both GCM matrix elements to shrink once the isoscalar pairing amplitude
is included as a generator coordinate \cite{Hinohara14,Menendez2015}.

\section{Summary}
\label{summary}

We have used covariant multi-reference density functional theory to treat
low-lying positive- and negative-parity states in $^{150}$Nd and $^{150}$Sm.
The GCM mixes symmetry-projected states with different amounts of quadrupole and
octupole deformation.  The results indicate that octupole shape correlation have
important dynamical effects, including the reduction of quadrupole collectivity
in the low-lying states of both nuclei.  As for $0\nu\beta\beta$ decay, static
deformation, whether quadrupole or octupole, quenches the nuclear matrix element
for that process, but shape fluctuations moderate the effect, so that the
inclusion of octupole degrees of freedom ends up reducing the matrix element
between the two nuclei considered here by only $7\%$.

\section*{Acknowledgements}

We are grateful to R. Rodr\'{\i}guez-Guzm\'{a}n, C. F. Jiao, and L. S. Song for
fruitful discussions and to T. R. Rodr\'{\i}guegz for providing us the
unpublished results of his non-relativistic GCM calculations.  Support for this
work was provided through the Scientific Discovery through Advanced Computing
(SciDAC) program funded by US Department of Energy, Office of Science, Advanced
Scientific Computing Research and Nuclear Physics, under Contract No.
DE-SC0008641, ER41896, and by the National Natural Science Foundation of China
under Grant Nos. 11575148, 11475140, and 11305134.

  \bibliographystyle{apsrev4-1}

\begin{thebibliography}{74}%
\makeatletter
\providecommand \@ifxundefined [1]{%
 \@ifx{#1\undefined}
}%
\providecommand \@ifnum [1]{%
 \ifnum #1\expandafter \@firstoftwo
 \else \expandafter \@secondoftwo
 \fi
}%
\providecommand \@ifx [1]{%
 \ifx #1\expandafter \@firstoftwo
 \else \expandafter \@secondoftwo
 \fi
}%
\providecommand \natexlab [1]{#1}%
\providecommand \enquote  [1]{``#1''}%
\providecommand \bibnamefont  [1]{#1}%
\providecommand \bibfnamefont [1]{#1}%
\providecommand \citenamefont [1]{#1}%
\providecommand \href@noop [0]{\@secondoftwo}%
\providecommand \href [0]{\begingroup \@sanitize@url \@href}%
\providecommand \@href[1]{\@@startlink{#1}\@@href}%
\providecommand \@@href[1]{\endgroup#1\@@endlink}%
\providecommand \@sanitize@url [0]{\catcode `\\12\catcode `\$12\catcode
  `\&12\catcode `\#12\catcode `\^12\catcode `\_12\catcode `\%12\relax}%
\providecommand \@@startlink[1]{}%
\providecommand \@@endlink[0]{}%
\providecommand \url  [0]{\begingroup\@sanitize@url \@url }%
\providecommand \@url [1]{\endgroup\@href {#1}{\urlprefix }}%
\providecommand \urlprefix  [0]{URL }%
\providecommand \Eprint [0]{\href }%
\providecommand \doibase [0]{http://dx.doi.org/}%
\providecommand \selectlanguage [0]{\@gobble}%
\providecommand \bibinfo  [0]{\@secondoftwo}%
\providecommand \bibfield  [0]{\@secondoftwo}%
\providecommand \translation [1]{[#1]}%
\providecommand \BibitemOpen [0]{}%
\providecommand \bibitemStop [0]{}%
\providecommand \bibitemNoStop [0]{.\EOS\space}%
\providecommand \EOS [0]{\spacefactor3000\relax}%
\providecommand \BibitemShut  [1]{\csname bibitem#1\endcsname}%
\let\auto@bib@innerbib\@empty
\bibitem [{\citenamefont {Argyriades}\ \emph {et~al.}(2009)\citenamefont
  {Argyriades}, \citenamefont {Arnold}, \citenamefont {Augier}, \citenamefont
  {Baker}, \citenamefont {Barabash}, \citenamefont {Basharina-Freshville},
  \citenamefont {Bongrand}, \citenamefont {Broudin}, \citenamefont {Brudanin},
  \citenamefont {Caffrey}, \citenamefont {Chauveau}, \citenamefont
  {Daraktchieva}, \citenamefont {Durand}, \citenamefont {Egorov}, \citenamefont
  {Fatemi-Ghomi}, \citenamefont {Flack}, \citenamefont {Hubert}, \citenamefont
  {Jerie}, \citenamefont {Jullian}, \citenamefont {Kauer}, \citenamefont
  {King}, \citenamefont {Klimenko}, \citenamefont {Kochetov}, \citenamefont
  {Konovalov}, \citenamefont {Kovalenko}, \citenamefont {Lalanne},
  \citenamefont {Lamhamdi}, \citenamefont {Lang}, \citenamefont {Lemi\`ere},
  \citenamefont {Longuemare}, \citenamefont {Lutter}, \citenamefont {Marquet},
  \citenamefont {Martin-Albo}, \citenamefont {Mauger}, \citenamefont {Nachab},
  \citenamefont {Nasteva}, \citenamefont {Nemchenok}, \citenamefont {Nova},
  \citenamefont {Novella}, \citenamefont {Ohsumi}, \citenamefont {Pahlka},
  \citenamefont {Perrot}, \citenamefont {Piquemal}, \citenamefont {Reyss},
  \citenamefont {Ricol}, \citenamefont {Saakyan}, \citenamefont {Sarazin},
  \citenamefont {Simard}, \citenamefont {\ifmmode~\check{S}\else
  \v{S}\fi{}imkovic}, \citenamefont {Shitov}, \citenamefont {Smolnikov},
  \citenamefont {Snow}, \citenamefont {S\"oldner-Rembold}, \citenamefont
  {\ifmmode~\check{S}\else \v{S}\fi{}tekl}, \citenamefont {Suhonen},
  \citenamefont {Sutton}, \citenamefont {Szklarz}, \citenamefont {Thomas},
  \citenamefont {Timkin}, \citenamefont {Tretyak}, \citenamefont {Umatov},
  \citenamefont {V\'ala}, \citenamefont {Vanyushin}, \citenamefont {Vasiliev},
  \citenamefont {Vorobel},\ and\ \citenamefont {Vylov}}]{Argyriades09}%
  \BibitemOpen
  \bibfield  {author} {\bibinfo {author} {\bibfnamefont {J.}~\bibnamefont
  {Argyriades}}, \bibinfo {author} {\bibfnamefont {R.}~\bibnamefont {Arnold}},
  \bibinfo {author} {\bibfnamefont {C.}~\bibnamefont {Augier}}, \bibinfo
  {author} {\bibfnamefont {J.}~\bibnamefont {Baker}}, \bibinfo {author}
  {\bibfnamefont {A.~S.}\ \bibnamefont {Barabash}}, \bibinfo {author}
  {\bibfnamefont {A.}~\bibnamefont {Basharina-Freshville}}, \bibinfo {author}
  {\bibfnamefont {M.}~\bibnamefont {Bongrand}}, \bibinfo {author}
  {\bibfnamefont {G.}~\bibnamefont {Broudin}}, \bibinfo {author} {\bibfnamefont
  {V.}~\bibnamefont {Brudanin}}, \bibinfo {author} {\bibfnamefont {A.~J.}\
  \bibnamefont {Caffrey}}, \bibinfo {author} {\bibfnamefont {E.}~\bibnamefont
  {Chauveau}}, \bibinfo {author} {\bibfnamefont {Z.}~\bibnamefont
  {Daraktchieva}}, \bibinfo {author} {\bibfnamefont {D.}~\bibnamefont
  {Durand}}, \bibinfo {author} {\bibfnamefont {V.}~\bibnamefont {Egorov}},
  \bibinfo {author} {\bibfnamefont {N.}~\bibnamefont {Fatemi-Ghomi}}, \bibinfo
  {author} {\bibfnamefont {R.}~\bibnamefont {Flack}}, \bibinfo {author}
  {\bibfnamefont {P.}~\bibnamefont {Hubert}}, \bibinfo {author} {\bibfnamefont
  {J.}~\bibnamefont {Jerie}}, \bibinfo {author} {\bibfnamefont
  {S.}~\bibnamefont {Jullian}}, \bibinfo {author} {\bibfnamefont
  {M.}~\bibnamefont {Kauer}}, \bibinfo {author} {\bibfnamefont
  {S.}~\bibnamefont {King}}, \bibinfo {author} {\bibfnamefont {A.}~\bibnamefont
  {Klimenko}}, \bibinfo {author} {\bibfnamefont {O.}~\bibnamefont {Kochetov}},
  \bibinfo {author} {\bibfnamefont {S.~I.}\ \bibnamefont {Konovalov}}, \bibinfo
  {author} {\bibfnamefont {V.}~\bibnamefont {Kovalenko}}, \bibinfo {author}
  {\bibfnamefont {D.}~\bibnamefont {Lalanne}}, \bibinfo {author} {\bibfnamefont
  {T.}~\bibnamefont {Lamhamdi}}, \bibinfo {author} {\bibfnamefont
  {K.}~\bibnamefont {Lang}}, \bibinfo {author} {\bibfnamefont {Y.}~\bibnamefont
  {Lemi\`ere}}, \bibinfo {author} {\bibfnamefont {C.}~\bibnamefont
  {Longuemare}}, \bibinfo {author} {\bibfnamefont {G.}~\bibnamefont {Lutter}},
  \bibinfo {author} {\bibfnamefont {C.}~\bibnamefont {Marquet}}, \bibinfo
  {author} {\bibfnamefont {J.}~\bibnamefont {Martin-Albo}}, \bibinfo {author}
  {\bibfnamefont {F.}~\bibnamefont {Mauger}}, \bibinfo {author} {\bibfnamefont
  {A.}~\bibnamefont {Nachab}}, \bibinfo {author} {\bibfnamefont
  {I.}~\bibnamefont {Nasteva}}, \bibinfo {author} {\bibfnamefont
  {I.}~\bibnamefont {Nemchenok}}, \bibinfo {author} {\bibfnamefont
  {F.}~\bibnamefont {Nova}}, \bibinfo {author} {\bibfnamefont {P.}~\bibnamefont
  {Novella}}, \bibinfo {author} {\bibfnamefont {H.}~\bibnamefont {Ohsumi}},
  \bibinfo {author} {\bibfnamefont {R.~B.}\ \bibnamefont {Pahlka}}, \bibinfo
  {author} {\bibfnamefont {F.}~\bibnamefont {Perrot}}, \bibinfo {author}
  {\bibfnamefont {F.}~\bibnamefont {Piquemal}}, \bibinfo {author}
  {\bibfnamefont {J.~L.}\ \bibnamefont {Reyss}}, \bibinfo {author}
  {\bibfnamefont {J.~S.}\ \bibnamefont {Ricol}}, \bibinfo {author}
  {\bibfnamefont {R.}~\bibnamefont {Saakyan}}, \bibinfo {author} {\bibfnamefont
  {X.}~\bibnamefont {Sarazin}}, \bibinfo {author} {\bibfnamefont
  {L.}~\bibnamefont {Simard}}, \bibinfo {author} {\bibfnamefont
  {F.}~\bibnamefont {\ifmmode~\check{S}\else \v{S}\fi{}imkovic}}, \bibinfo
  {author} {\bibfnamefont {Y.}~\bibnamefont {Shitov}}, \bibinfo {author}
  {\bibfnamefont {A.}~\bibnamefont {Smolnikov}}, \bibinfo {author}
  {\bibfnamefont {S.}~\bibnamefont {Snow}}, \bibinfo {author} {\bibfnamefont
  {S.}~\bibnamefont {S\"oldner-Rembold}}, \bibinfo {author} {\bibfnamefont
  {I.}~\bibnamefont {\ifmmode~\check{S}\else \v{S}\fi{}tekl}}, \bibinfo
  {author} {\bibfnamefont {J.}~\bibnamefont {Suhonen}}, \bibinfo {author}
  {\bibfnamefont {C.~S.}\ \bibnamefont {Sutton}}, \bibinfo {author}
  {\bibfnamefont {G.}~\bibnamefont {Szklarz}}, \bibinfo {author} {\bibfnamefont
  {J.}~\bibnamefont {Thomas}}, \bibinfo {author} {\bibfnamefont
  {V.}~\bibnamefont {Timkin}}, \bibinfo {author} {\bibfnamefont
  {V.}~\bibnamefont {Tretyak}}, \bibinfo {author} {\bibfnamefont
  {V.}~\bibnamefont {Umatov}}, \bibinfo {author} {\bibfnamefont
  {L.}~\bibnamefont {V\'ala}}, \bibinfo {author} {\bibfnamefont
  {I.}~\bibnamefont {Vanyushin}}, \bibinfo {author} {\bibfnamefont
  {V.}~\bibnamefont {Vasiliev}}, \bibinfo {author} {\bibfnamefont
  {V.}~\bibnamefont {Vorobel}}, \ and\ \bibinfo {author} {\bibfnamefont
  {T.}~\bibnamefont {Vylov}} (\bibinfo {collaboration} {NEMO Collaboration}),\
  }\href {\doibase 10.1103/PhysRevC.80.032501} {\bibfield  {journal} {\bibinfo
  {journal} {Phys. Rev. C}\ }\textbf {\bibinfo {volume} {80}},\ \bibinfo
  {pages} {032501} (\bibinfo {year} {2009})}\BibitemShut {NoStop}%
\bibitem [{\citenamefont {Bongrand}(2015)}]{Bongrand2015}%
  \BibitemOpen
  \bibfield  {author} {\bibinfo {author} {\bibfnamefont {M.}~\bibnamefont
  {Bongrand}},\ }\href {\doibase http://dx.doi.org/10.1016/j.phpro.2014.12.034}
  {\bibfield  {journal} {\bibinfo  {journal} {Physics Procedia}\ }\textbf
  {\bibinfo {volume} {61}},\ \bibinfo {pages} {211 } (\bibinfo {year}
  {2015})},\ \bibinfo {note} {13th International Conference on Topics in
  Astroparticle and Underground Physics, \{TAUP\} 2013}\BibitemShut {NoStop}%
\bibitem [{\citenamefont {Casten}\ and\ \citenamefont
  {Zamfir}(2001)}]{Casten2001}%
  \BibitemOpen
  \bibfield  {author} {\bibinfo {author} {\bibfnamefont {R.~F.}\ \bibnamefont
  {Casten}}\ and\ \bibinfo {author} {\bibfnamefont {N.~V.}\ \bibnamefont
  {Zamfir}},\ }\href {\doibase 10.1103/PhysRevLett.87.052503} {\bibfield
  {journal} {\bibinfo  {journal} {Phys. Rev. Lett.}\ }\textbf {\bibinfo
  {volume} {87}},\ \bibinfo {pages} {052503} (\bibinfo {year}
  {2001})}\BibitemShut {NoStop}%
\bibitem [{\citenamefont {Kr\"ucken}\ \emph {et~al.}(2002)\citenamefont
  {Kr\"ucken}, \citenamefont {Albanna}, \citenamefont {Bialik}, \citenamefont
  {Casten}, \citenamefont {Cooper}, \citenamefont {Dewald}, \citenamefont
  {Zamfir}, \citenamefont {Barton}, \citenamefont {Beausang}, \citenamefont
  {Caprio}, \citenamefont {Hecht}, \citenamefont {Klug}, \citenamefont {Novak},
  \citenamefont {Pietralla},\ and\ \citenamefont {von Brentano}}]{Krucken2002}%
  \BibitemOpen
  \bibfield  {author} {\bibinfo {author} {\bibfnamefont {R.}~\bibnamefont
  {Kr\"ucken}}, \bibinfo {author} {\bibfnamefont {B.}~\bibnamefont {Albanna}},
  \bibinfo {author} {\bibfnamefont {C.}~\bibnamefont {Bialik}}, \bibinfo
  {author} {\bibfnamefont {R.~F.}\ \bibnamefont {Casten}}, \bibinfo {author}
  {\bibfnamefont {J.~R.}\ \bibnamefont {Cooper}}, \bibinfo {author}
  {\bibfnamefont {A.}~\bibnamefont {Dewald}}, \bibinfo {author} {\bibfnamefont
  {N.~V.}\ \bibnamefont {Zamfir}}, \bibinfo {author} {\bibfnamefont {C.~J.}\
  \bibnamefont {Barton}}, \bibinfo {author} {\bibfnamefont {C.~W.}\
  \bibnamefont {Beausang}}, \bibinfo {author} {\bibfnamefont {M.~A.}\
  \bibnamefont {Caprio}}, \bibinfo {author} {\bibfnamefont {A.~A.}\
  \bibnamefont {Hecht}}, \bibinfo {author} {\bibfnamefont {T.}~\bibnamefont
  {Klug}}, \bibinfo {author} {\bibfnamefont {J.~R.}\ \bibnamefont {Novak}},
  \bibinfo {author} {\bibfnamefont {N.}~\bibnamefont {Pietralla}}, \ and\
  \bibinfo {author} {\bibfnamefont {P.}~\bibnamefont {von Brentano}},\ }\href
  {\doibase 10.1103/PhysRevLett.88.232501} {\bibfield  {journal} {\bibinfo
  {journal} {Phys. Rev. Lett.}\ }\textbf {\bibinfo {volume} {88}},\ \bibinfo
  {pages} {232501} (\bibinfo {year} {2002})}\BibitemShut {NoStop}%
\bibitem [{\citenamefont {Meng}\ \emph {et~al.}(2005)\citenamefont {Meng},
  \citenamefont {Zhang}, \citenamefont {Zhou}, \citenamefont {Toki},\ and\
  \citenamefont {Geng}}]{Meng05}%
  \BibitemOpen
  \bibfield  {author} {\bibinfo {author} {\bibfnamefont {J.}~\bibnamefont
  {Meng}}, \bibinfo {author} {\bibfnamefont {W.}~\bibnamefont {Zhang}},
  \bibinfo {author} {\bibfnamefont {S.~G.}\ \bibnamefont {Zhou}}, \bibinfo
  {author} {\bibfnamefont {H.}~\bibnamefont {Toki}}, \ and\ \bibinfo {author}
  {\bibfnamefont {L.~S.}\ \bibnamefont {Geng}},\ }\href {\doibase
  10.1140/epja/i2005-10066-6} {\bibfield  {journal} {\bibinfo  {journal} {Euro.
  Phys. J. A}\ }\textbf {\bibinfo {volume} {25}},\ \bibinfo {pages} {23}
  (\bibinfo {year} {2005})}\BibitemShut {NoStop}%
\bibitem [{\citenamefont {Sheng}\ and\ \citenamefont {Guo}(2005)}]{Sheng05}%
  \BibitemOpen
  \bibfield  {author} {\bibinfo {author} {\bibfnamefont {Z.~Q.}\ \bibnamefont
  {Sheng}}\ and\ \bibinfo {author} {\bibfnamefont {J.-Y.}\ \bibnamefont
  {Guo}},\ }\href {\doibase 10.1142/S0217732305017883} {\bibfield  {journal}
  {\bibinfo  {journal} {Mod. Phys. Lett. A}\ }\textbf {\bibinfo {volume}
  {20}},\ \bibinfo {pages} {2711} (\bibinfo {year} {2005})}\BibitemShut
  {NoStop}%
\bibitem [{\citenamefont {Nik\ifmmode \check{s}\else
  \v{s}\fi{}i\ifmmode~\acute{c}\else \'{c}\fi{}}\ \emph
  {et~al.}(2007)\citenamefont {Nik\ifmmode \check{s}\else
  \v{s}\fi{}i\ifmmode~\acute{c}\else \'{c}\fi{}}, \citenamefont {Vretenar},
  \citenamefont {Lalazissis},\ and\ \citenamefont {Ring}}]{Niksic07}%
  \BibitemOpen
  \bibfield  {author} {\bibinfo {author} {\bibfnamefont {T.}~\bibnamefont
  {Nik\ifmmode \check{s}\else \v{s}\fi{}i\ifmmode~\acute{c}\else \'{c}\fi{}}},
  \bibinfo {author} {\bibfnamefont {D.}~\bibnamefont {Vretenar}}, \bibinfo
  {author} {\bibfnamefont {G.~A.}\ \bibnamefont {Lalazissis}}, \ and\ \bibinfo
  {author} {\bibfnamefont {P.}~\bibnamefont {Ring}},\ }\href {\doibase
  10.1103/PhysRevLett.99.092502} {\bibfield  {journal} {\bibinfo  {journal}
  {Phys. Rev. Lett.}\ }\textbf {\bibinfo {volume} {99}},\ \bibinfo {pages}
  {092502} (\bibinfo {year} {2007})}\BibitemShut {NoStop}%
\bibitem [{\citenamefont {Rodr\'{\i}guez-Guzm\'an}\ and\ \citenamefont
  {Sarriguren}(2007)}]{Guzman07}%
  \BibitemOpen
  \bibfield  {author} {\bibinfo {author} {\bibfnamefont {R.}~\bibnamefont
  {Rodr\'{\i}guez-Guzm\'an}}\ and\ \bibinfo {author} {\bibfnamefont
  {P.}~\bibnamefont {Sarriguren}},\ }\href {\doibase
  10.1103/PhysRevC.76.064303} {\bibfield  {journal} {\bibinfo  {journal} {Phys.
  Rev. C}\ }\textbf {\bibinfo {volume} {76}},\ \bibinfo {pages} {064303}
  (\bibinfo {year} {2007})}\BibitemShut {NoStop}%
\bibitem [{\citenamefont {Rodriguez}\ and\ \citenamefont
  {Egido}(2008)}]{Rodriguez08}%
  \BibitemOpen
  \bibfield  {author} {\bibinfo {author} {\bibfnamefont {T.~R.}\ \bibnamefont
  {Rodriguez}}\ and\ \bibinfo {author} {\bibfnamefont {J.~L.}\ \bibnamefont
  {Egido}},\ }\href {\doibase http://dx.doi.org/10.1016/j.physletb.2008.03.061}
  {\bibfield  {journal} {\bibinfo  {journal} {Phys. Lett. B}\ }\textbf
  {\bibinfo {volume} {663}},\ \bibinfo {pages} {49 } (\bibinfo {year}
  {2008})}\BibitemShut {NoStop}%
\bibitem [{\citenamefont {Robledo}\ \emph {et~al.}(2008)\citenamefont
  {Robledo}, \citenamefont {Rodr\'{\i}guez-Guzm\'an},\ and\ \citenamefont
  {Sarriguren}}]{Robledo08}%
  \BibitemOpen
  \bibfield  {author} {\bibinfo {author} {\bibfnamefont {L.~M.}\ \bibnamefont
  {Robledo}}, \bibinfo {author} {\bibfnamefont {R.~R.}\ \bibnamefont
  {Rodr\'{\i}guez-Guzm\'an}}, \ and\ \bibinfo {author} {\bibfnamefont
  {P.}~\bibnamefont {Sarriguren}},\ }\href {\doibase
  10.1103/PhysRevC.78.034314} {\bibfield  {journal} {\bibinfo  {journal} {Phys.
  Rev. C}\ }\textbf {\bibinfo {volume} {78}},\ \bibinfo {pages} {034314}
  (\bibinfo {year} {2008})}\BibitemShut {NoStop}%
\bibitem [{\citenamefont {Li}\ \emph {et~al.}(2009)\citenamefont {Li},
  \citenamefont {Nik\ifmmode \check{s}\else \v{s}\fi{}i\ifmmode~\acute{c}\else
  \'{c}\fi{}}, \citenamefont {Vretenar}, \citenamefont {Meng}, \citenamefont
  {Lalazissis},\ and\ \citenamefont {Ring}}]{Li09}%
  \BibitemOpen
  \bibfield  {author} {\bibinfo {author} {\bibfnamefont {Z.~P.}\ \bibnamefont
  {Li}}, \bibinfo {author} {\bibfnamefont {T.}~\bibnamefont {Nik\ifmmode
  \check{s}\else \v{s}\fi{}i\ifmmode~\acute{c}\else \'{c}\fi{}}}, \bibinfo
  {author} {\bibfnamefont {D.}~\bibnamefont {Vretenar}}, \bibinfo {author}
  {\bibfnamefont {J.}~\bibnamefont {Meng}}, \bibinfo {author} {\bibfnamefont
  {G.~A.}\ \bibnamefont {Lalazissis}}, \ and\ \bibinfo {author} {\bibfnamefont
  {P.}~\bibnamefont {Ring}},\ }\href {\doibase 10.1103/PhysRevC.79.054301}
  {\bibfield  {journal} {\bibinfo  {journal} {Phys. Rev. C}\ }\textbf {\bibinfo
  {volume} {79}},\ \bibinfo {pages} {054301} (\bibinfo {year}
  {2009})}\BibitemShut {NoStop}%
\bibitem [{\citenamefont {Urban}\ \emph {et~al.}(1987)\citenamefont {Urban},
  \citenamefont {Lieder}, \citenamefont {Gast}, \citenamefont {Hebbinghaus},
  \citenamefont {Kr?mer-Flecken}, \citenamefont {Blume},\ and\ \citenamefont
  {H¨¹bel}}]{Urban87}%
  \BibitemOpen
  \bibfield  {author} {\bibinfo {author} {\bibfnamefont {W.}~\bibnamefont
  {Urban}}, \bibinfo {author} {\bibfnamefont {R.}~\bibnamefont {Lieder}},
  \bibinfo {author} {\bibfnamefont {W.}~\bibnamefont {Gast}}, \bibinfo {author}
  {\bibfnamefont {G.}~\bibnamefont {Hebbinghaus}}, \bibinfo {author}
  {\bibfnamefont {A.}~\bibnamefont {Kr?mer-Flecken}}, \bibinfo {author}
  {\bibfnamefont {K.}~\bibnamefont {Blume}}, \ and\ \bibinfo {author}
  {\bibfnamefont {H.}~\bibnamefont {H¨¹bel}},\ }\href {\doibase
  http://dx.doi.org/10.1016/0370-2693(87)91009-4} {\bibfield  {journal}
  {\bibinfo  {journal} {Phys. Lett. B}\ }\textbf {\bibinfo {volume} {185}},\
  \bibinfo {pages} {331 } (\bibinfo {year} {1987})}\BibitemShut {NoStop}%
\bibitem [{\citenamefont {Friedrichs}\ \emph {et~al.}(1992)\citenamefont
  {Friedrichs}, \citenamefont {Schlitt}, \citenamefont {Margraf}, \citenamefont
  {Lindenstruth}, \citenamefont {Wesselborg}, \citenamefont {Heil},
  \citenamefont {Pitz}, \citenamefont {Kneissl}, \citenamefont {von Brentano},
  \citenamefont {Herzberg}, \citenamefont {Zilges}, \citenamefont {H\"ager},
  \citenamefont {M\"uller},\ and\ \citenamefont {Schumacher}}]{Friedrichs92}%
  \BibitemOpen
  \bibfield  {author} {\bibinfo {author} {\bibfnamefont {H.}~\bibnamefont
  {Friedrichs}}, \bibinfo {author} {\bibfnamefont {B.}~\bibnamefont {Schlitt}},
  \bibinfo {author} {\bibfnamefont {J.}~\bibnamefont {Margraf}}, \bibinfo
  {author} {\bibfnamefont {S.}~\bibnamefont {Lindenstruth}}, \bibinfo {author}
  {\bibfnamefont {C.}~\bibnamefont {Wesselborg}}, \bibinfo {author}
  {\bibfnamefont {R.~D.}\ \bibnamefont {Heil}}, \bibinfo {author}
  {\bibfnamefont {H.~H.}\ \bibnamefont {Pitz}}, \bibinfo {author}
  {\bibfnamefont {U.}~\bibnamefont {Kneissl}}, \bibinfo {author} {\bibfnamefont
  {P.}~\bibnamefont {von Brentano}}, \bibinfo {author} {\bibfnamefont {R.~D.}\
  \bibnamefont {Herzberg}}, \bibinfo {author} {\bibfnamefont {A.}~\bibnamefont
  {Zilges}}, \bibinfo {author} {\bibfnamefont {D.}~\bibnamefont {H\"ager}},
  \bibinfo {author} {\bibfnamefont {G.}~\bibnamefont {M\"uller}}, \ and\
  \bibinfo {author} {\bibfnamefont {M.}~\bibnamefont {Schumacher}},\ }\href
  {\doibase 10.1103/PhysRevC.45.R892} {\bibfield  {journal} {\bibinfo
  {journal} {Phys. Rev. C}\ }\textbf {\bibinfo {volume} {45}},\ \bibinfo
  {pages} {R892} (\bibinfo {year} {1992})}\BibitemShut {NoStop}%
\bibitem [{\citenamefont {Bvumbi}\ \emph {et~al.}(2013)\citenamefont {Bvumbi},
  \citenamefont {Sharpey-Schafer}, \citenamefont {Jones}, \citenamefont
  {Mullins}, \citenamefont {Nyak\'o}, \citenamefont {Juh\'asz}, \citenamefont
  {Bark}, \citenamefont {Bianco}, \citenamefont {Cullen}, \citenamefont
  {Curien}, \citenamefont {Garrett}, \citenamefont {Greenlees}, \citenamefont
  {Hirvonen}, \citenamefont {Jakobsson}, \citenamefont {Kau}, \citenamefont
  {Komati}, \citenamefont {Julin}, \citenamefont {Juutinen}, \citenamefont
  {Ketelhut}, \citenamefont {Korichi}, \citenamefont {Lawrie}, \citenamefont
  {Lawrie}, \citenamefont {Leino}, \citenamefont {Madiba}, \citenamefont
  {Majola}, \citenamefont {Maine}, \citenamefont {Minkova}, \citenamefont
  {Ncapayi}, \citenamefont {Nieminen}, \citenamefont {Peura}, \citenamefont
  {Rahkila}, \citenamefont {Riedinger}, \citenamefont {Ruotsalainen},
  \citenamefont {Saren}, \citenamefont {Scholey}, \citenamefont {Sorri},
  \citenamefont {Stolze}, \citenamefont {Timar}, \citenamefont {Uusitalo},\
  and\ \citenamefont {Vymers}}]{Bvumbi13}%
  \BibitemOpen
  \bibfield  {author} {\bibinfo {author} {\bibfnamefont {S.~P.}\ \bibnamefont
  {Bvumbi}}, \bibinfo {author} {\bibfnamefont {J.~F.}\ \bibnamefont
  {Sharpey-Schafer}}, \bibinfo {author} {\bibfnamefont {P.~M.}\ \bibnamefont
  {Jones}}, \bibinfo {author} {\bibfnamefont {S.~M.}\ \bibnamefont {Mullins}},
  \bibinfo {author} {\bibfnamefont {B.~M.}\ \bibnamefont {Nyak\'o}}, \bibinfo
  {author} {\bibfnamefont {K.}~\bibnamefont {Juh\'asz}}, \bibinfo {author}
  {\bibfnamefont {R.~A.}\ \bibnamefont {Bark}}, \bibinfo {author}
  {\bibfnamefont {L.}~\bibnamefont {Bianco}}, \bibinfo {author} {\bibfnamefont
  {D.~M.}\ \bibnamefont {Cullen}}, \bibinfo {author} {\bibfnamefont
  {D.}~\bibnamefont {Curien}}, \bibinfo {author} {\bibfnamefont {P.~E.}\
  \bibnamefont {Garrett}}, \bibinfo {author} {\bibfnamefont {P.~T.}\
  \bibnamefont {Greenlees}}, \bibinfo {author} {\bibfnamefont {J.}~\bibnamefont
  {Hirvonen}}, \bibinfo {author} {\bibfnamefont {U.}~\bibnamefont {Jakobsson}},
  \bibinfo {author} {\bibfnamefont {J.}~\bibnamefont {Kau}}, \bibinfo {author}
  {\bibfnamefont {F.}~\bibnamefont {Komati}}, \bibinfo {author} {\bibfnamefont
  {R.}~\bibnamefont {Julin}}, \bibinfo {author} {\bibfnamefont
  {S.}~\bibnamefont {Juutinen}}, \bibinfo {author} {\bibfnamefont
  {S.}~\bibnamefont {Ketelhut}}, \bibinfo {author} {\bibfnamefont
  {A.}~\bibnamefont {Korichi}}, \bibinfo {author} {\bibfnamefont {E.~A.}\
  \bibnamefont {Lawrie}}, \bibinfo {author} {\bibfnamefont {J.~J.}\
  \bibnamefont {Lawrie}}, \bibinfo {author} {\bibfnamefont {M.}~\bibnamefont
  {Leino}}, \bibinfo {author} {\bibfnamefont {T.~E.}\ \bibnamefont {Madiba}},
  \bibinfo {author} {\bibfnamefont {S.~N.~T.}\ \bibnamefont {Majola}}, \bibinfo
  {author} {\bibfnamefont {P.}~\bibnamefont {Maine}}, \bibinfo {author}
  {\bibfnamefont {A.}~\bibnamefont {Minkova}}, \bibinfo {author} {\bibfnamefont
  {N.~J.}\ \bibnamefont {Ncapayi}}, \bibinfo {author} {\bibfnamefont
  {P.}~\bibnamefont {Nieminen}}, \bibinfo {author} {\bibfnamefont
  {P.}~\bibnamefont {Peura}}, \bibinfo {author} {\bibfnamefont
  {P.}~\bibnamefont {Rahkila}}, \bibinfo {author} {\bibfnamefont {L.~L.}\
  \bibnamefont {Riedinger}}, \bibinfo {author} {\bibfnamefont {P.}~\bibnamefont
  {Ruotsalainen}}, \bibinfo {author} {\bibfnamefont {J.}~\bibnamefont {Saren}},
  \bibinfo {author} {\bibfnamefont {C.}~\bibnamefont {Scholey}}, \bibinfo
  {author} {\bibfnamefont {J.}~\bibnamefont {Sorri}}, \bibinfo {author}
  {\bibfnamefont {S.}~\bibnamefont {Stolze}}, \bibinfo {author} {\bibfnamefont
  {J.}~\bibnamefont {Timar}}, \bibinfo {author} {\bibfnamefont
  {J.}~\bibnamefont {Uusitalo}}, \ and\ \bibinfo {author} {\bibfnamefont
  {P.~A.}\ \bibnamefont {Vymers}},\ }\href {\doibase
  10.1103/PhysRevC.87.044333} {\bibfield  {journal} {\bibinfo  {journal} {Phys.
  Rev. C}\ }\textbf {\bibinfo {volume} {87}},\ \bibinfo {pages} {044333}
  (\bibinfo {year} {2013})}\BibitemShut {NoStop}%
\bibitem [{\citenamefont {Nazarewicz}\ and\ \citenamefont
  {Tabor}(1992)}]{Nazarewicz92}%
  \BibitemOpen
  \bibfield  {author} {\bibinfo {author} {\bibfnamefont {W.}~\bibnamefont
  {Nazarewicz}}\ and\ \bibinfo {author} {\bibfnamefont {S.~L.}\ \bibnamefont
  {Tabor}},\ }\href {\doibase 10.1103/PhysRevC.45.2226} {\bibfield  {journal}
  {\bibinfo  {journal} {Phys. Rev. C}\ }\textbf {\bibinfo {volume} {45}},\
  \bibinfo {pages} {2226} (\bibinfo {year} {1992})}\BibitemShut {NoStop}%
\bibitem [{\citenamefont {Garrote}\ \emph {et~al.}(1998)\citenamefont
  {Garrote}, \citenamefont {Egido},\ and\ \citenamefont {Robledo}}]{Garrote98}%
  \BibitemOpen
  \bibfield  {author} {\bibinfo {author} {\bibfnamefont {E.}~\bibnamefont
  {Garrote}}, \bibinfo {author} {\bibfnamefont {J.~L.}\ \bibnamefont {Egido}},
  \ and\ \bibinfo {author} {\bibfnamefont {L.~M.}\ \bibnamefont {Robledo}},\
  }\href {\doibase 10.1103/PhysRevLett.80.4398} {\bibfield  {journal} {\bibinfo
   {journal} {Phys. Rev. Lett.}\ }\textbf {\bibinfo {volume} {80}},\ \bibinfo
  {pages} {4398} (\bibinfo {year} {1998})}\BibitemShut {NoStop}%
\bibitem [{\citenamefont {Babilon}\ \emph {et~al.}(2005)\citenamefont
  {Babilon}, \citenamefont {Zamfir}, \citenamefont {Kusnezov}, \citenamefont
  {McCutchan},\ and\ \citenamefont {Zilges}}]{Babilon05}%
  \BibitemOpen
  \bibfield  {author} {\bibinfo {author} {\bibfnamefont {M.}~\bibnamefont
  {Babilon}}, \bibinfo {author} {\bibfnamefont {N.~V.}\ \bibnamefont {Zamfir}},
  \bibinfo {author} {\bibfnamefont {D.}~\bibnamefont {Kusnezov}}, \bibinfo
  {author} {\bibfnamefont {E.~A.}\ \bibnamefont {McCutchan}}, \ and\ \bibinfo
  {author} {\bibfnamefont {A.}~\bibnamefont {Zilges}},\ }\href {\doibase
  10.1103/PhysRevC.72.064302} {\bibfield  {journal} {\bibinfo  {journal} {Phys.
  Rev. C}\ }\textbf {\bibinfo {volume} {72}},\ \bibinfo {pages} {064302}
  (\bibinfo {year} {2005})}\BibitemShut {NoStop}%
\bibitem [{\citenamefont {Minkov}\ \emph {et~al.}(2006)\citenamefont {Minkov},
  \citenamefont {Yotov}, \citenamefont {Drenska}, \citenamefont {Scheid},
  \citenamefont {Bonatsos}, \citenamefont {Lenis},\ and\ \citenamefont
  {Petrellis}}]{Minkov06}%
  \BibitemOpen
  \bibfield  {author} {\bibinfo {author} {\bibfnamefont {N.}~\bibnamefont
  {Minkov}}, \bibinfo {author} {\bibfnamefont {P.}~\bibnamefont {Yotov}},
  \bibinfo {author} {\bibfnamefont {S.}~\bibnamefont {Drenska}}, \bibinfo
  {author} {\bibfnamefont {W.}~\bibnamefont {Scheid}}, \bibinfo {author}
  {\bibfnamefont {D.}~\bibnamefont {Bonatsos}}, \bibinfo {author}
  {\bibfnamefont {D.}~\bibnamefont {Lenis}}, \ and\ \bibinfo {author}
  {\bibfnamefont {D.}~\bibnamefont {Petrellis}},\ }\href {\doibase
  10.1103/PhysRevC.73.044315} {\bibfield  {journal} {\bibinfo  {journal} {Phys.
  Rev. C}\ }\textbf {\bibinfo {volume} {73}},\ \bibinfo {pages} {044315}
  (\bibinfo {year} {2006})}\BibitemShut {NoStop}%
\bibitem [{\citenamefont {Zhang}\ \emph {et~al.}(2010)\citenamefont {Zhang},
  \citenamefont {Li}, \citenamefont {Zhang},\ and\ \citenamefont
  {Meng}}]{Zhang10}%
  \BibitemOpen
  \bibfield  {author} {\bibinfo {author} {\bibfnamefont {W.}~\bibnamefont
  {Zhang}}, \bibinfo {author} {\bibfnamefont {Z.~P.}\ \bibnamefont {Li}},
  \bibinfo {author} {\bibfnamefont {S.~Q.}\ \bibnamefont {Zhang}}, \ and\
  \bibinfo {author} {\bibfnamefont {J.}~\bibnamefont {Meng}},\ }\href {\doibase
  10.1103/PhysRevC.81.034302} {\bibfield  {journal} {\bibinfo  {journal} {Phys.
  Rev. C}\ }\textbf {\bibinfo {volume} {81}},\ \bibinfo {pages} {034302}
  (\bibinfo {year} {2010})}\BibitemShut {NoStop}%
\bibitem [{\citenamefont {Robledo}\ and\ \citenamefont
  {Bertsch}(2011)}]{Robledo11}%
  \BibitemOpen
  \bibfield  {author} {\bibinfo {author} {\bibfnamefont {L.~M.}\ \bibnamefont
  {Robledo}}\ and\ \bibinfo {author} {\bibfnamefont {G.~F.}\ \bibnamefont
  {Bertsch}},\ }\href {\doibase 10.1103/PhysRevC.84.054302} {\bibfield
  {journal} {\bibinfo  {journal} {Phys. Rev. C}\ }\textbf {\bibinfo {volume}
  {84}},\ \bibinfo {pages} {054302} (\bibinfo {year} {2011})}\BibitemShut
  {NoStop}%
\bibitem [{\citenamefont {Rodr\'{\i}guez-Guzm\'an}\ \emph
  {et~al.}(2012)\citenamefont {Rodr\'{\i}guez-Guzm\'an}, \citenamefont
  {Robledo},\ and\ \citenamefont {Sarriguren}}]{Guzman2012}%
  \BibitemOpen
  \bibfield  {author} {\bibinfo {author} {\bibfnamefont {R.}~\bibnamefont
  {Rodr\'{\i}guez-Guzm\'an}}, \bibinfo {author} {\bibfnamefont {L.~M.}\
  \bibnamefont {Robledo}}, \ and\ \bibinfo {author} {\bibfnamefont
  {P.}~\bibnamefont {Sarriguren}},\ }\href {\doibase
  10.1103/PhysRevC.86.034336} {\bibfield  {journal} {\bibinfo  {journal} {Phys.
  Rev. C}\ }\textbf {\bibinfo {volume} {86}},\ \bibinfo {pages} {034336}
  (\bibinfo {year} {2012})}\BibitemShut {NoStop}%
\bibitem [{\citenamefont {Nomura}\ \emph {et~al.}(2014)\citenamefont {Nomura},
  \citenamefont {Vretenar}, \citenamefont {Nik\ifmmode \check{s}\else
  \v{s}\fi{}i\ifmmode~\acute{c}\else \'{c}\fi{}},\ and\ \citenamefont
  {Lu}}]{Nomura2014}%
  \BibitemOpen
  \bibfield  {author} {\bibinfo {author} {\bibfnamefont {K.}~\bibnamefont
  {Nomura}}, \bibinfo {author} {\bibfnamefont {D.}~\bibnamefont {Vretenar}},
  \bibinfo {author} {\bibfnamefont {T.}~\bibnamefont {Nik\ifmmode
  \check{s}\else \v{s}\fi{}i\ifmmode~\acute{c}\else \'{c}\fi{}}}, \ and\
  \bibinfo {author} {\bibfnamefont {B.-N.}\ \bibnamefont {Lu}},\ }\href
  {\doibase 10.1103/PhysRevC.89.024312} {\bibfield  {journal} {\bibinfo
  {journal} {Phys. Rev. C}\ }\textbf {\bibinfo {volume} {89}},\ \bibinfo
  {pages} {024312} (\bibinfo {year} {2014})}\BibitemShut {NoStop}%
\bibitem [{\citenamefont {Nomura}\ \emph {et~al.}(2015)\citenamefont {Nomura},
  \citenamefont {Rodr\'{\i}guez-Guzm\'an},\ and\ \citenamefont
  {Robledo}}]{Nomura2015}%
  \BibitemOpen
  \bibfield  {author} {\bibinfo {author} {\bibfnamefont {K.}~\bibnamefont
  {Nomura}}, \bibinfo {author} {\bibfnamefont {R.}~\bibnamefont
  {Rodr\'{\i}guez-Guzm\'an}}, \ and\ \bibinfo {author} {\bibfnamefont {L.~M.}\
  \bibnamefont {Robledo}},\ }\href {\doibase 10.1103/PhysRevC.92.014312}
  {\bibfield  {journal} {\bibinfo  {journal} {Phys. Rev. C}\ }\textbf {\bibinfo
  {volume} {92}},\ \bibinfo {pages} {014312} (\bibinfo {year}
  {2015})}\BibitemShut {NoStop}%
\bibitem [{\citenamefont {Reinhard}(1989)}]{Reinhard89}%
  \BibitemOpen
  \bibfield  {author} {\bibinfo {author} {\bibfnamefont {P.~G.}\ \bibnamefont
  {Reinhard}},\ }\href {http://stacks.iop.org/0034-4885/52/i=4/a=002}
  {\bibfield  {journal} {\bibinfo  {journal} {Rep. Prog. Phys.}\ }\textbf
  {\bibinfo {volume} {52}},\ \bibinfo {pages} {439} (\bibinfo {year}
  {1989})}\BibitemShut {NoStop}%
\bibitem [{\citenamefont {Ring}(1996)}]{Ring96}%
  \BibitemOpen
  \bibfield  {author} {\bibinfo {author} {\bibfnamefont {P.}~\bibnamefont
  {Ring}},\ }\href {\doibase http://dx.doi.org/10.1016/0146-6410(96)00054-3}
  {\bibfield  {journal} {\bibinfo  {journal} {Prog. Part. Nucl. Phys.}\
  }\textbf {\bibinfo {volume} {37}},\ \bibinfo {pages} {193 } (\bibinfo {year}
  {1996})}\BibitemShut {NoStop}%
\bibitem [{\citenamefont {Vretenar}\ \emph {et~al.}(2005)\citenamefont
  {Vretenar}, \citenamefont {Afanasjev}, \citenamefont {Lalazissis},\ and\
  \citenamefont {Ring}}]{Vretenar05}%
  \BibitemOpen
  \bibfield  {author} {\bibinfo {author} {\bibfnamefont {D.}~\bibnamefont
  {Vretenar}}, \bibinfo {author} {\bibfnamefont {A.}~\bibnamefont {Afanasjev}},
  \bibinfo {author} {\bibfnamefont {G.}~\bibnamefont {Lalazissis}}, \ and\
  \bibinfo {author} {\bibfnamefont {P.}~\bibnamefont {Ring}},\ }\href {\doibase
  http://dx.doi.org/10.1016/j.physrep.2004.10.001} {\bibfield  {journal}
  {\bibinfo  {journal} {Phys. Rep.}\ }\textbf {\bibinfo {volume} {409}},\
  \bibinfo {pages} {101 } (\bibinfo {year} {2005})}\BibitemShut {NoStop}%
\bibitem [{\citenamefont {Meng}\ \emph {et~al.}(2006)\citenamefont {Meng},
  \citenamefont {Toki}, \citenamefont {Zhou}, \citenamefont {Zhang},
  \citenamefont {Long},\ and\ \citenamefont {Geng}}]{Meng06}%
  \BibitemOpen
  \bibfield  {author} {\bibinfo {author} {\bibfnamefont {J.}~\bibnamefont
  {Meng}}, \bibinfo {author} {\bibfnamefont {H.}~\bibnamefont {Toki}}, \bibinfo
  {author} {\bibfnamefont {S.}~\bibnamefont {Zhou}}, \bibinfo {author}
  {\bibfnamefont {S.}~\bibnamefont {Zhang}}, \bibinfo {author} {\bibfnamefont
  {W.}~\bibnamefont {Long}}, \ and\ \bibinfo {author} {\bibfnamefont
  {L.}~\bibnamefont {Geng}},\ }\href {\doibase
  http://dx.doi.org/10.1016/j.ppnp.2005.06.001} {\bibfield  {journal} {\bibinfo
   {journal} {Prog. Part. Nucl. Phys.}\ }\textbf {\bibinfo {volume} {57}},\
  \bibinfo {pages} {470 } (\bibinfo {year} {2006})}\BibitemShut {NoStop}%
\bibitem [{\citenamefont {Geng}\ \emph {et~al.}(2007)\citenamefont {Geng},
  \citenamefont {Meng},\ and\ \citenamefont {Toki}}]{Geng07}%
  \BibitemOpen
  \bibfield  {author} {\bibinfo {author} {\bibfnamefont {L.~S.}\ \bibnamefont
  {Geng}}, \bibinfo {author} {\bibfnamefont {J.}~\bibnamefont {Meng}}, \ and\
  \bibinfo {author} {\bibfnamefont {H.}~\bibnamefont {Toki}},\ }\href
  {http://stacks.iop.org/0256-307X/24/i=7/a=021} {\bibfield  {journal}
  {\bibinfo  {journal} {Chin. Phys. Lett.}\ }\textbf {\bibinfo {volume} {24}},\
  \bibinfo {pages} {1865} (\bibinfo {year} {2007})}\BibitemShut {NoStop}%
\bibitem [{\citenamefont {Guo}\ \emph {et~al.}(2010)\citenamefont {Guo},
  \citenamefont {Jiao},\ and\ \citenamefont {Fang}}]{Guo10}%
  \BibitemOpen
  \bibfield  {author} {\bibinfo {author} {\bibfnamefont {J.-Y.}\ \bibnamefont
  {Guo}}, \bibinfo {author} {\bibfnamefont {P.}~\bibnamefont {Jiao}}, \ and\
  \bibinfo {author} {\bibfnamefont {X.-Z.}\ \bibnamefont {Fang}},\ }\href
  {\doibase 10.1103/PhysRevC.82.047301} {\bibfield  {journal} {\bibinfo
  {journal} {Phys. Rev. C}\ }\textbf {\bibinfo {volume} {82}},\ \bibinfo
  {pages} {047301} (\bibinfo {year} {2010})}\BibitemShut {NoStop}%
\bibitem [{\citenamefont {Lu}\ \emph {et~al.}(2012)\citenamefont {Lu},
  \citenamefont {Zhao},\ and\ \citenamefont {Zhou}}]{Lu12}%
  \BibitemOpen
  \bibfield  {author} {\bibinfo {author} {\bibfnamefont {B.-N.}\ \bibnamefont
  {Lu}}, \bibinfo {author} {\bibfnamefont {E.-G.}\ \bibnamefont {Zhao}}, \ and\
  \bibinfo {author} {\bibfnamefont {S.-G.}\ \bibnamefont {Zhou}},\ }\href
  {\doibase 10.1103/PhysRevC.85.011301} {\bibfield  {journal} {\bibinfo
  {journal} {Phys. Rev. C}\ }\textbf {\bibinfo {volume} {85}},\ \bibinfo
  {pages} {011301} (\bibinfo {year} {2012})}\BibitemShut {NoStop}%
\bibitem [{\citenamefont {Li}\ \emph {et~al.}(2013)\citenamefont {Li},
  \citenamefont {Song}, \citenamefont {Yao}, \citenamefont {Vretenar},\ and\
  \citenamefont {Meng}}]{Li13}%
  \BibitemOpen
  \bibfield  {author} {\bibinfo {author} {\bibfnamefont {Z.~P.}\ \bibnamefont
  {Li}}, \bibinfo {author} {\bibfnamefont {B.~Y.}\ \bibnamefont {Song}},
  \bibinfo {author} {\bibfnamefont {J.~M.}\ \bibnamefont {Yao}}, \bibinfo
  {author} {\bibfnamefont {D.}~\bibnamefont {Vretenar}}, \ and\ \bibinfo
  {author} {\bibfnamefont {J.}~\bibnamefont {Meng}},\ }\href {\doibase
  http://dx.doi.org/10.1016/j.physletb.2013.09.035} {\bibfield  {journal}
  {\bibinfo  {journal} {Phys. Lett. B}\ }\textbf {\bibinfo {volume} {726}},\
  \bibinfo {pages} {866 } (\bibinfo {year} {2013})}\BibitemShut {NoStop}%
\bibitem [{\citenamefont {Nik\ifmmode \check{s}\else
  \v{s}\fi{}i\ifmmode~\acute{c}\else \'{c}\fi{}}\ \emph
  {et~al.}(2006)\citenamefont {Nik\ifmmode \check{s}\else
  \v{s}\fi{}i\ifmmode~\acute{c}\else \'{c}\fi{}}, \citenamefont {Vretenar},\
  and\ \citenamefont {Ring}}]{Niksic06}%
  \BibitemOpen
  \bibfield  {author} {\bibinfo {author} {\bibfnamefont {T.}~\bibnamefont
  {Nik\ifmmode \check{s}\else \v{s}\fi{}i\ifmmode~\acute{c}\else \'{c}\fi{}}},
  \bibinfo {author} {\bibfnamefont {D.}~\bibnamefont {Vretenar}}, \ and\
  \bibinfo {author} {\bibfnamefont {P.}~\bibnamefont {Ring}},\ }\href {\doibase
  10.1103/PhysRevC.74.064309} {\bibfield  {journal} {\bibinfo  {journal} {Phys.
  Rev. C}\ }\textbf {\bibinfo {volume} {74}},\ \bibinfo {pages} {064309}
  (\bibinfo {year} {2006})}\BibitemShut {NoStop}%
\bibitem [{\citenamefont {Yao}\ \emph {et~al.}(2013)\citenamefont {Yao},
  \citenamefont {Mei},\ and\ \citenamefont {Li}}]{Yao13}%
  \BibitemOpen
  \bibfield  {author} {\bibinfo {author} {\bibfnamefont {J.~M.}\ \bibnamefont
  {Yao}}, \bibinfo {author} {\bibfnamefont {H.}~\bibnamefont {Mei}}, \ and\
  \bibinfo {author} {\bibfnamefont {Z.}~\bibnamefont {Li}},\ }\href {\doibase
  http://dx.doi.org/10.1016/j.physletb.2013.05.049} {\bibfield  {journal}
  {\bibinfo  {journal} {Phys. Lett. B}\ }\textbf {\bibinfo {volume} {723}},\
  \bibinfo {pages} {459 } (\bibinfo {year} {2013})}\BibitemShut {NoStop}%
\bibitem [{\citenamefont {Wu}\ \emph {et~al.}(2014)\citenamefont {Wu},
  \citenamefont {Yao},\ and\ \citenamefont {Li}}]{Wu14}%
  \BibitemOpen
  \bibfield  {author} {\bibinfo {author} {\bibfnamefont {X.~Y.}\ \bibnamefont
  {Wu}}, \bibinfo {author} {\bibfnamefont {J.~M.}\ \bibnamefont {Yao}}, \ and\
  \bibinfo {author} {\bibfnamefont {Z.~P.}\ \bibnamefont {Li}},\ }\href
  {\doibase 10.1103/PhysRevC.89.017304} {\bibfield  {journal} {\bibinfo
  {journal} {Phys. Rev. C}\ }\textbf {\bibinfo {volume} {89}},\ \bibinfo
  {pages} {017304} (\bibinfo {year} {2014})}\BibitemShut {NoStop}%
\bibitem [{\citenamefont {Wu}\ and\ \citenamefont {Zhou}(2015)}]{Wu15}%
  \BibitemOpen
  \bibfield  {author} {\bibinfo {author} {\bibfnamefont {X.~Y.}\ \bibnamefont
  {Wu}}\ and\ \bibinfo {author} {\bibfnamefont {X.~R.}\ \bibnamefont {Zhou}},\
  }\href {\doibase 10.1103/PhysRevC.92.054321} {\bibfield  {journal} {\bibinfo
  {journal} {Phys. Rev. C}\ }\textbf {\bibinfo {volume} {92}},\ \bibinfo
  {pages} {054321} (\bibinfo {year} {2015})}\BibitemShut {NoStop}%
\bibitem [{\citenamefont {Yao}\ \emph {et~al.}(2010)\citenamefont {Yao},
  \citenamefont {Meng}, \citenamefont {Ring},\ and\ \citenamefont
  {Vretenar}}]{Yao10}%
  \BibitemOpen
  \bibfield  {author} {\bibinfo {author} {\bibfnamefont {J.~M.}\ \bibnamefont
  {Yao}}, \bibinfo {author} {\bibfnamefont {J.}~\bibnamefont {Meng}}, \bibinfo
  {author} {\bibfnamefont {P.}~\bibnamefont {Ring}}, \ and\ \bibinfo {author}
  {\bibfnamefont {D.}~\bibnamefont {Vretenar}},\ }\href {\doibase
  10.1103/PhysRevC.81.044311} {\bibfield  {journal} {\bibinfo  {journal} {Phys.
  Rev. C}\ }\textbf {\bibinfo {volume} {81}},\ \bibinfo {pages} {044311}
  (\bibinfo {year} {2010})}\BibitemShut {NoStop}%
\bibitem [{\citenamefont {Yao}\ \emph {et~al.}(2011)\citenamefont {Yao},
  \citenamefont {Mei}, \citenamefont {Chen}, \citenamefont {Meng},
  \citenamefont {Ring},\ and\ \citenamefont {Vretenar}}]{Yao11}%
  \BibitemOpen
  \bibfield  {author} {\bibinfo {author} {\bibfnamefont {J.~M.}\ \bibnamefont
  {Yao}}, \bibinfo {author} {\bibfnamefont {H.}~\bibnamefont {Mei}}, \bibinfo
  {author} {\bibfnamefont {H.}~\bibnamefont {Chen}}, \bibinfo {author}
  {\bibfnamefont {J.}~\bibnamefont {Meng}}, \bibinfo {author} {\bibfnamefont
  {P.}~\bibnamefont {Ring}}, \ and\ \bibinfo {author} {\bibfnamefont
  {D.}~\bibnamefont {Vretenar}},\ }\href {\doibase 10.1103/PhysRevC.83.014308}
  {\bibfield  {journal} {\bibinfo  {journal} {Phys. Rev. C}\ }\textbf {\bibinfo
  {volume} {83}},\ \bibinfo {pages} {014308} (\bibinfo {year}
  {2011})}\BibitemShut {NoStop}%
\bibitem [{\citenamefont {Yao}\ \emph {et~al.}(2014)\citenamefont {Yao},
  \citenamefont {Hagino}, \citenamefont {Li}, \citenamefont {Meng},\ and\
  \citenamefont {Ring}}]{Yao14}%
  \BibitemOpen
  \bibfield  {author} {\bibinfo {author} {\bibfnamefont {J.~M.}\ \bibnamefont
  {Yao}}, \bibinfo {author} {\bibfnamefont {K.}~\bibnamefont {Hagino}},
  \bibinfo {author} {\bibfnamefont {Z.~P.}\ \bibnamefont {Li}}, \bibinfo
  {author} {\bibfnamefont {J.}~\bibnamefont {Meng}}, \ and\ \bibinfo {author}
  {\bibfnamefont {P.}~\bibnamefont {Ring}},\ }\href {\doibase
  10.1103/PhysRevC.89.054306} {\bibfield  {journal} {\bibinfo  {journal} {Phys.
  Rev. C}\ }\textbf {\bibinfo {volume} {89}},\ \bibinfo {pages} {054306}
  (\bibinfo {year} {2014})}\BibitemShut {NoStop}%
\bibitem [{\citenamefont {Yao}\ \emph {et~al.}(2015{\natexlab{a}})\citenamefont
  {Yao}, \citenamefont {Zhou},\ and\ \citenamefont {Li}}]{Yao15-Ra}%
  \BibitemOpen
  \bibfield  {author} {\bibinfo {author} {\bibfnamefont {J.~M.}\ \bibnamefont
  {Yao}}, \bibinfo {author} {\bibfnamefont {E.~F.}\ \bibnamefont {Zhou}}, \
  and\ \bibinfo {author} {\bibfnamefont {Z.~P.}\ \bibnamefont {Li}},\ }\href
  {\doibase 10.1103/PhysRevC.92.041304} {\bibfield  {journal} {\bibinfo
  {journal} {Phys. Rev. C}\ }\textbf {\bibinfo {volume} {92}},\ \bibinfo
  {pages} {041304} (\bibinfo {year} {2015}{\natexlab{a}})}\BibitemShut
  {NoStop}%
\bibitem [{\citenamefont {Zhou}\ \emph {et~al.}(2016)\citenamefont {Zhou},
  \citenamefont {Yao}, \citenamefont {Li}, \citenamefont {Meng},\ and\
  \citenamefont {Ring}}]{Zhou16}%
  \BibitemOpen
  \bibfield  {author} {\bibinfo {author} {\bibfnamefont {E.~F.}\ \bibnamefont
  {Zhou}}, \bibinfo {author} {\bibfnamefont {J.~M.}\ \bibnamefont {Yao}},
  \bibinfo {author} {\bibfnamefont {Z.~P.}\ \bibnamefont {Li}}, \bibinfo
  {author} {\bibfnamefont {J.}~\bibnamefont {Meng}}, \ and\ \bibinfo {author}
  {\bibfnamefont {P.}~\bibnamefont {Ring}},\ }\href {\doibase
  http://dx.doi.org/10.1016/j.physletb.2015.12.028} {\bibfield  {journal}
  {\bibinfo  {journal} {Phys. Lett. B}\ }\textbf {\bibinfo {volume} {753}},\
  \bibinfo {pages} {227 } (\bibinfo {year} {2016})}\BibitemShut {NoStop}%
\bibitem [{\citenamefont {Hirsch}\ \emph {et~al.}(1995)\citenamefont {Hirsch},
  \citenamefont {Castanos},\ and\ \citenamefont {Hess}}]{Hirsch95}%
  \BibitemOpen
  \bibfield  {author} {\bibinfo {author} {\bibfnamefont {J.~G.}\ \bibnamefont
  {Hirsch}}, \bibinfo {author} {\bibfnamefont {O.}~\bibnamefont {Castanos}}, \
  and\ \bibinfo {author} {\bibfnamefont {P.~O.}\ \bibnamefont {Hess}},\ }\href
  {\doibase http://dx.doi.org/10.1016/0375-9474(94)00464-X} {\bibfield
  {journal} {\bibinfo  {journal} {Nucl. Phys. A}\ }\textbf {\bibinfo {volume}
  {582}},\ \bibinfo {pages} {124} (\bibinfo {year} {1995})}\BibitemShut
  {NoStop}%
\bibitem [{\citenamefont {Chaturvedi}\ \emph {et~al.}(2008)\citenamefont
  {Chaturvedi}, \citenamefont {Chandra}, \citenamefont {Rath}, \citenamefont
  {Raina},\ and\ \citenamefont {Hirsch}}]{Chaturvedi2008}%
  \BibitemOpen
  \bibfield  {author} {\bibinfo {author} {\bibfnamefont {K.}~\bibnamefont
  {Chaturvedi}}, \bibinfo {author} {\bibfnamefont {R.}~\bibnamefont {Chandra}},
  \bibinfo {author} {\bibfnamefont {P.~K.}\ \bibnamefont {Rath}}, \bibinfo
  {author} {\bibfnamefont {P.~K.}\ \bibnamefont {Raina}}, \ and\ \bibinfo
  {author} {\bibfnamefont {J.~G.}\ \bibnamefont {Hirsch}},\ }\href
  {http://dx.doi.org/10.1103/PhysRevC.78.054302} {\bibfield  {journal}
  {\bibinfo  {journal} {Phys. Rev. C}\ }\textbf {\bibinfo {volume} {78}},\
  \bibinfo {pages} {054302} (\bibinfo {year} {2008})}\BibitemShut {NoStop}%
\bibitem [{\citenamefont {Fang}\ \emph {et~al.}(2010)\citenamefont {Fang},
  \citenamefont {Faessler}, \citenamefont {Rodin},\ and\ \citenamefont
  {\ifmmode~\check{S}\else \v{S}\fi{}imkovic}}]{Fang2010}%
  \BibitemOpen
  \bibfield  {author} {\bibinfo {author} {\bibfnamefont {D.-L.}\ \bibnamefont
  {Fang}}, \bibinfo {author} {\bibfnamefont {A.}~\bibnamefont {Faessler}},
  \bibinfo {author} {\bibfnamefont {V.}~\bibnamefont {Rodin}}, \ and\ \bibinfo
  {author} {\bibfnamefont {F.}~\bibnamefont {\ifmmode~\check{S}\else
  \v{S}\fi{}imkovic}},\ }\href {\doibase 10.1103/PhysRevC.82.051301} {\bibfield
   {journal} {\bibinfo  {journal} {Phys. Rev. C}\ }\textbf {\bibinfo {volume}
  {82}},\ \bibinfo {pages} {051301} (\bibinfo {year} {2010})}\BibitemShut
  {NoStop}%
\bibitem [{\citenamefont {Rodr{\'\i}guez}\ and\ \citenamefont
  {Martinez-Pinedo}(2010)}]{Rodriguez2010}%
  \BibitemOpen
  \bibfield  {author} {\bibinfo {author} {\bibfnamefont {T.~R.}\ \bibnamefont
  {Rodr{\'\i}guez}}\ and\ \bibinfo {author} {\bibfnamefont {G.}~\bibnamefont
  {Martinez-Pinedo}},\ }\href
  {http://dx.doi.org/10.1103/PhysRevLett.105.252503} {\bibfield  {journal}
  {\bibinfo  {journal} {Phys. Rev. Lett.}\ }\textbf {\bibinfo {volume} {105}},\
  \bibinfo {pages} {252503} (\bibinfo {year} {2010})}\BibitemShut {NoStop}%
\bibitem [{\citenamefont {Mustonen}\ and\ \citenamefont
  {Engel}(2013)}]{Mustonen2013}%
  \BibitemOpen
  \bibfield  {author} {\bibinfo {author} {\bibfnamefont {M.~T.}\ \bibnamefont
  {Mustonen}}\ and\ \bibinfo {author} {\bibfnamefont {J.}~\bibnamefont
  {Engel}},\ }\href {http://dx.doi.org/10.1103/PhysRevC.87.064302} {\bibfield
  {journal} {\bibinfo  {journal} {Phys. Rev. C}\ }\textbf {\bibinfo {volume}
  {87}},\ \bibinfo {pages} {064302} (\bibinfo {year} {2013})}\BibitemShut
  {NoStop}%
\bibitem [{\citenamefont {Song}\ \emph {et~al.}(2014)\citenamefont {Song},
  \citenamefont {Yao}, \citenamefont {Ring},\ and\ \citenamefont
  {Meng}}]{Song2014}%
  \BibitemOpen
  \bibfield  {author} {\bibinfo {author} {\bibfnamefont {L.~S.}\ \bibnamefont
  {Song}}, \bibinfo {author} {\bibfnamefont {J.~M.}\ \bibnamefont {Yao}},
  \bibinfo {author} {\bibfnamefont {P.}~\bibnamefont {Ring}}, \ and\ \bibinfo
  {author} {\bibfnamefont {J.}~\bibnamefont {Meng}},\ }\href
  {http://dx.doi.org/10.1103/PhysRevC.90.054309} {\bibfield  {journal}
  {\bibinfo  {journal} {Phys. Rev. C}\ }\textbf {\bibinfo {volume} {90}},\
  \bibinfo {pages} {054309} (\bibinfo {year} {2014})}\BibitemShut {NoStop}%
\bibitem [{\citenamefont {Terasaki}(2015)}]{Terasaki2014}%
  \BibitemOpen
  \bibfield  {author} {\bibinfo {author} {\bibfnamefont {J.}~\bibnamefont
  {Terasaki}},\ }\href {\doibase 10.1103/PhysRevC.91.034318} {\bibfield
  {journal} {\bibinfo  {journal} {Phys. Rev. C}\ }\textbf {\bibinfo {volume}
  {91}},\ \bibinfo {pages} {034318} (\bibinfo {year} {2015})}\BibitemShut
  {NoStop}%
\bibitem [{\citenamefont {Yao}\ \emph {et~al.}(2015{\natexlab{b}})\citenamefont
  {Yao}, \citenamefont {Song}, \citenamefont {Hagino}, \citenamefont {Ring},\
  and\ \citenamefont {Meng}}]{Yao2015}%
  \BibitemOpen
  \bibfield  {author} {\bibinfo {author} {\bibfnamefont {J.~M.}\ \bibnamefont
  {Yao}}, \bibinfo {author} {\bibfnamefont {L.~S.}\ \bibnamefont {Song}},
  \bibinfo {author} {\bibfnamefont {K.}~\bibnamefont {Hagino}}, \bibinfo
  {author} {\bibfnamefont {P.}~\bibnamefont {Ring}}, \ and\ \bibinfo {author}
  {\bibfnamefont {J.}~\bibnamefont {Meng}},\ }\href
  {http://dx.doi.org/10.1103/PhysRevC.91.024316} {\bibfield  {journal}
  {\bibinfo  {journal} {Phys. Rev. C}\ }\textbf {\bibinfo {volume} {91}},\
  \bibinfo {pages} {024316} (\bibinfo {year} {2015}{\natexlab{b}})}\BibitemShut
  {NoStop}%
\bibitem [{\citenamefont {Nikolaus}\ \emph {et~al.}(1992)\citenamefont
  {Nikolaus}, \citenamefont {Hoch},\ and\ \citenamefont
  {Madland}}]{Nikolaus92}%
  \BibitemOpen
  \bibfield  {author} {\bibinfo {author} {\bibfnamefont {B.~A.}\ \bibnamefont
  {Nikolaus}}, \bibinfo {author} {\bibfnamefont {T.}~\bibnamefont {Hoch}}, \
  and\ \bibinfo {author} {\bibfnamefont {D.~G.}\ \bibnamefont {Madland}},\
  }\href {\doibase 10.1103/PhysRevC.46.1757} {\bibfield  {journal} {\bibinfo
  {journal} {Phys. Rev. C}\ }\textbf {\bibinfo {volume} {46}},\ \bibinfo
  {pages} {1757} (\bibinfo {year} {1992})}\BibitemShut {NoStop}%
\bibitem [{\citenamefont {B\"urvenich}\ \emph {et~al.}(2002)\citenamefont
  {B\"urvenich}, \citenamefont {Madland}, \citenamefont {Maruhn},\ and\
  \citenamefont {Reinhard}}]{Burvenich02}%
  \BibitemOpen
  \bibfield  {author} {\bibinfo {author} {\bibfnamefont {T.}~\bibnamefont
  {B\"urvenich}}, \bibinfo {author} {\bibfnamefont {D.~G.}\ \bibnamefont
  {Madland}}, \bibinfo {author} {\bibfnamefont {J.~A.}\ \bibnamefont {Maruhn}},
  \ and\ \bibinfo {author} {\bibfnamefont {P.-G.}\ \bibnamefont {Reinhard}},\
  }\href {\doibase 10.1103/PhysRevC.65.044308} {\bibfield  {journal} {\bibinfo
  {journal} {Phys. Rev. C}\ }\textbf {\bibinfo {volume} {65}},\ \bibinfo
  {pages} {044308} (\bibinfo {year} {2002})}\BibitemShut {NoStop}%
\bibitem [{\citenamefont {Zhao}\ \emph {et~al.}(2010)\citenamefont {Zhao},
  \citenamefont {Li}, \citenamefont {Yao},\ and\ \citenamefont
  {Meng}}]{Zhao10}%
  \BibitemOpen
  \bibfield  {author} {\bibinfo {author} {\bibfnamefont {P.~W.}\ \bibnamefont
  {Zhao}}, \bibinfo {author} {\bibfnamefont {Z.~P.}\ \bibnamefont {Li}},
  \bibinfo {author} {\bibfnamefont {J.~M.}\ \bibnamefont {Yao}}, \ and\
  \bibinfo {author} {\bibfnamefont {J.}~\bibnamefont {Meng}},\ }\href {\doibase
  10.1103/PhysRevC.82.054319} {\bibfield  {journal} {\bibinfo  {journal} {Phys.
  Rev. C}\ }\textbf {\bibinfo {volume} {82}},\ \bibinfo {pages} {054319}
  (\bibinfo {year} {2010})}\BibitemShut {NoStop}%
\bibitem [{\citenamefont {Gambhir}\ \emph {et~al.}(1990)\citenamefont
  {Gambhir}, \citenamefont {Ring},\ and\ \citenamefont {Thimet}}]{Gambhir90}%
  \BibitemOpen
  \bibfield  {author} {\bibinfo {author} {\bibfnamefont {Y.~K.}\ \bibnamefont
  {Gambhir}}, \bibinfo {author} {\bibfnamefont {P.}~\bibnamefont {Ring}}, \
  and\ \bibinfo {author} {\bibfnamefont {A.}~\bibnamefont {Thimet}},\ }\href
  {\doibase http://dx.doi.org/10.1016/0003-4916(90)90330-Q} {\bibfield
  {journal} {\bibinfo  {journal} {Annals of Physics}\ }\textbf {\bibinfo
  {volume} {198}},\ \bibinfo {pages} {132 } (\bibinfo {year}
  {1990})}\BibitemShut {NoStop}%
\bibitem [{\citenamefont {Krieger}\ \emph {et~al.}(1990)\citenamefont
  {Krieger}, \citenamefont {Bonche}, \citenamefont {Flocard}, \citenamefont
  {Quentin},\ and\ \citenamefont {Weiss}}]{Krieger90}%
  \BibitemOpen
  \bibfield  {author} {\bibinfo {author} {\bibfnamefont {S.~J.}\ \bibnamefont
  {Krieger}}, \bibinfo {author} {\bibfnamefont {P.}~\bibnamefont {Bonche}},
  \bibinfo {author} {\bibfnamefont {H.}~\bibnamefont {Flocard}}, \bibinfo
  {author} {\bibfnamefont {P.}~\bibnamefont {Quentin}}, \ and\ \bibinfo
  {author} {\bibfnamefont {M.}~\bibnamefont {Weiss}},\ }\href {\doibase
  http://dx.doi.org/10.1016/0375-9474(90)90035-K} {\bibfield  {journal}
  {\bibinfo  {journal} {Nucl. Phys. A}\ }\textbf {\bibinfo {volume} {517}},\
  \bibinfo {pages} {275 } (\bibinfo {year} {1990})}\BibitemShut {NoStop}%
\bibitem [{\citenamefont {Li}\ \emph {et~al.}(2011)\citenamefont {Li},
  \citenamefont {Xiang}, \citenamefont {Yao}, \citenamefont {Chen},\ and\
  \citenamefont {Meng}}]{Li11}%
  \BibitemOpen
  \bibfield  {author} {\bibinfo {author} {\bibfnamefont {Z.}~\bibnamefont
  {Li}}, \bibinfo {author} {\bibfnamefont {J.}~\bibnamefont {Xiang}}, \bibinfo
  {author} {\bibfnamefont {J.~M.}\ \bibnamefont {Yao}}, \bibinfo {author}
  {\bibfnamefont {H.}~\bibnamefont {Chen}}, \ and\ \bibinfo {author}
  {\bibfnamefont {J.}~\bibnamefont {Meng}},\ }\href {\doibase
  http://dx.doi.org/10.1142/S0218301311017909} {\bibfield  {journal} {\bibinfo
  {journal} {Int. J. Mod. Phys. E}\ }\textbf {\bibinfo {volume} {20}},\
  \bibinfo {pages} {494} (\bibinfo {year} {2011})}\BibitemShut {NoStop}%
\bibitem [{\citenamefont {Tian}\ \emph {et~al.}(2009)\citenamefont {Tian},
  \citenamefont {Ma},\ and\ \citenamefont {Ring}}]{Tian2009}%
  \BibitemOpen
  \bibfield  {author} {\bibinfo {author} {\bibfnamefont {Y.}~\bibnamefont
  {Tian}}, \bibinfo {author} {\bibfnamefont {Z.}~\bibnamefont {Ma}}, \ and\
  \bibinfo {author} {\bibfnamefont {P.}~\bibnamefont {Ring}},\ }\href {\doibase
  http://dx.doi.org/10.1016/j.physletb.2009.04.067} {\bibfield  {journal}
  {\bibinfo  {journal} {Phys. Lett. B}\ }\textbf {\bibinfo {volume} {676}},\
  \bibinfo {pages} {44 } (\bibinfo {year} {2009})}\BibitemShut {NoStop}%
\bibitem [{\citenamefont {Ring}\ and\ \citenamefont {Schuck}(1980)}]{Ring1980}%
  \BibitemOpen
  \bibfield  {author} {\bibinfo {author} {\bibfnamefont {P.}~\bibnamefont
  {Ring}}\ and\ \bibinfo {author} {\bibfnamefont {P.}~\bibnamefont {Schuck}},\
  }\href@noop {} {\emph {\bibinfo {title} {The Nuclear Many-Body Problem}}}\
  (\bibinfo  {publisher} {Springer-Verlag},\ \bibinfo {year}
  {1980})\BibitemShut {NoStop}%
\bibitem [{\citenamefont {Hill}\ and\ \citenamefont {Wheeler}(1953)}]{Hill53}%
  \BibitemOpen
  \bibfield  {author} {\bibinfo {author} {\bibfnamefont {D.~L.}\ \bibnamefont
  {Hill}}\ and\ \bibinfo {author} {\bibfnamefont {J.~A.}\ \bibnamefont
  {Wheeler}},\ }\href {\doibase 10.1103/PhysRev.89.1102} {\bibfield  {journal}
  {\bibinfo  {journal} {Phys. Rev.}\ }\textbf {\bibinfo {volume} {89}},\
  \bibinfo {pages} {1102} (\bibinfo {year} {1953})}\BibitemShut {NoStop}%
\bibitem [{\citenamefont {Griffin}\ and\ \citenamefont
  {Wheeler}(1957)}]{Griffin57}%
  \BibitemOpen
  \bibfield  {author} {\bibinfo {author} {\bibfnamefont {J.~J.}\ \bibnamefont
  {Griffin}}\ and\ \bibinfo {author} {\bibfnamefont {J.~A.}\ \bibnamefont
  {Wheeler}},\ }\href {\doibase 10.1103/PhysRev.108.311} {\bibfield  {journal}
  {\bibinfo  {journal} {Phys. Rev.}\ }\textbf {\bibinfo {volume} {108}},\
  \bibinfo {pages} {311} (\bibinfo {year} {1957})}\BibitemShut {NoStop}%
\bibitem [{\citenamefont {Rodriguez-Guzman}\ \emph {et~al.}(2002)\citenamefont
  {Rodriguez-Guzman}, \citenamefont {Egido},\ and\ \citenamefont
  {Robledo}}]{Guzman2002}%
  \BibitemOpen
  \bibfield  {author} {\bibinfo {author} {\bibfnamefont {R.}~\bibnamefont
  {Rodriguez-Guzman}}, \bibinfo {author} {\bibfnamefont {J.}~\bibnamefont
  {Egido}}, \ and\ \bibinfo {author} {\bibfnamefont {L.}~\bibnamefont
  {Robledo}},\ }\href {\doibase
  http://dx.doi.org/10.1016/S0375-9474(02)01019-9} {\bibfield  {journal}
  {\bibinfo  {journal} {Nucl. Phys. A}\ }\textbf {\bibinfo {volume} {709}},\
  \bibinfo {pages} {201 } (\bibinfo {year} {2002})}\BibitemShut {NoStop}%
\bibitem [{\citenamefont {Lacroix}\ \emph {et~al.}(2009)\citenamefont
  {Lacroix}, \citenamefont {Duguet},\ and\ \citenamefont {Bender}}]{Lacroix09}%
  \BibitemOpen
  \bibfield  {author} {\bibinfo {author} {\bibfnamefont {D.}~\bibnamefont
  {Lacroix}}, \bibinfo {author} {\bibfnamefont {T.}~\bibnamefont {Duguet}}, \
  and\ \bibinfo {author} {\bibfnamefont {M.}~\bibnamefont {Bender}},\ }\href
  {\doibase 10.1103/PhysRevC.79.044318} {\bibfield  {journal} {\bibinfo
  {journal} {Phys. Rev. C}\ }\textbf {\bibinfo {volume} {79}},\ \bibinfo
  {pages} {044318} (\bibinfo {year} {2009})}\BibitemShut {NoStop}%
\bibitem [{\citenamefont {Duguet}\ \emph {et~al.}(2009)\citenamefont {Duguet},
  \citenamefont {Bender}, \citenamefont {Bennaceur}, \citenamefont {Lacroix},\
  and\ \citenamefont {Lesinski}}]{Duguet09}%
  \BibitemOpen
  \bibfield  {author} {\bibinfo {author} {\bibfnamefont {T.}~\bibnamefont
  {Duguet}}, \bibinfo {author} {\bibfnamefont {M.}~\bibnamefont {Bender}},
  \bibinfo {author} {\bibfnamefont {K.}~\bibnamefont {Bennaceur}}, \bibinfo
  {author} {\bibfnamefont {D.}~\bibnamefont {Lacroix}}, \ and\ \bibinfo
  {author} {\bibfnamefont {T.}~\bibnamefont {Lesinski}},\ }\href {\doibase
  10.1103/PhysRevC.79.044320} {\bibfield  {journal} {\bibinfo  {journal} {Phys.
  Rev. C}\ }\textbf {\bibinfo {volume} {79}},\ \bibinfo {pages} {044320}
  (\bibinfo {year} {2009})}\BibitemShut {NoStop}%
\bibitem [{\citenamefont {Fomenko}(1970)}]{Fomenko1970}%
  \BibitemOpen
  \bibfield  {author} {\bibinfo {author} {\bibfnamefont {V.~N.}\ \bibnamefont
  {Fomenko}},\ }\href {\doibase 10.1088/0305-4470/3/1/002} {\bibfield
  {journal} {\bibinfo  {journal} {J. Phys. A: Gen. Phys.}\ }\textbf {\bibinfo
  {volume} {3}},\ \bibinfo {pages} {8 } (\bibinfo {year} {1970})}\BibitemShut
  {NoStop}%
\bibitem [{\citenamefont {Bender}\ and\ \citenamefont
  {Heenen}(2008)}]{Bender08}%
  \BibitemOpen
  \bibfield  {author} {\bibinfo {author} {\bibfnamefont {M.}~\bibnamefont
  {Bender}}\ and\ \bibinfo {author} {\bibfnamefont {P.-H.}\ \bibnamefont
  {Heenen}},\ }\href {\doibase 10.1103/PhysRevC.78.024309} {\bibfield
  {journal} {\bibinfo  {journal} {Phys. Rev. C}\ }\textbf {\bibinfo {volume}
  {78}},\ \bibinfo {pages} {024309} (\bibinfo {year} {2008})}\BibitemShut
  {NoStop}%
\bibitem [{\citenamefont {Rodr\'{\i}guez}\ and\ \citenamefont
  {Egido}(2010)}]{Rodriguez10}%
  \BibitemOpen
  \bibfield  {author} {\bibinfo {author} {\bibfnamefont {T.~R.}\ \bibnamefont
  {Rodr\'{\i}guez}}\ and\ \bibinfo {author} {\bibfnamefont {J.~L.}\
  \bibnamefont {Egido}},\ }\href {\doibase 10.1103/PhysRevC.81.064323}
  {\bibfield  {journal} {\bibinfo  {journal} {Phys. Rev. C}\ }\textbf {\bibinfo
  {volume} {81}},\ \bibinfo {pages} {064323} (\bibinfo {year}
  {2010})}\BibitemShut {NoStop}%
\bibitem [{\citenamefont {Avignone}\ \emph {et~al.}(2008)\citenamefont
  {Avignone}, \citenamefont {Elliott},\ and\ \citenamefont
  {Engel}}]{Avignone08}%
  \BibitemOpen
  \bibfield  {author} {\bibinfo {author} {\bibfnamefont {F.~T.}\ \bibnamefont
  {Avignone}}, \bibinfo {author} {\bibfnamefont {S.~R.}\ \bibnamefont
  {Elliott}}, \ and\ \bibinfo {author} {\bibfnamefont {J.}~\bibnamefont
  {Engel}},\ }\href {http://dx.doi.org/10.1103/RevModPhys.80.481} {\bibfield
  {journal} {\bibinfo  {journal} {Rev. Mod. Phys.}\ }\textbf {\bibinfo {volume}
  {80}},\ \bibinfo {pages} {481} (\bibinfo {year} {2008})}\BibitemShut
  {NoStop}%
\bibitem [{\citenamefont {Haxton}\ and\ \citenamefont
  {Stephenson}(1984)}]{Haxton84}%
  \BibitemOpen
  \bibfield  {author} {\bibinfo {author} {\bibfnamefont {W.~C.}\ \bibnamefont
  {Haxton}}\ and\ \bibinfo {author} {\bibfnamefont {G.~J.}\ \bibnamefont
  {Stephenson}},\ }\href {\doibase
  http://dx.doi.org/10.1016/0146-6410(84)90006-1} {\bibfield  {journal}
  {\bibinfo  {journal} {Prog. Part. Nucl. Phys.}\ }\textbf {\bibinfo {volume}
  {12}},\ \bibinfo {pages} {409 } (\bibinfo {year} {1984})}\BibitemShut
  {NoStop}%
\bibitem [{\citenamefont {\ifmmode~\check{S}\else \v{S}\fi{}imkovic}\ \emph
  {et~al.}(1999)\citenamefont {\ifmmode~\check{S}\else \v{S}\fi{}imkovic},
  \citenamefont {Pantis}, \citenamefont {Vergados},\ and\ \citenamefont
  {Faessler}}]{Simkovic1999}%
  \BibitemOpen
  \bibfield  {author} {\bibinfo {author} {\bibfnamefont {F.}~\bibnamefont
  {\ifmmode~\check{S}\else \v{S}\fi{}imkovic}}, \bibinfo {author}
  {\bibfnamefont {G.}~\bibnamefont {Pantis}}, \bibinfo {author} {\bibfnamefont
  {J.~D.}\ \bibnamefont {Vergados}}, \ and\ \bibinfo {author} {\bibfnamefont
  {A.}~\bibnamefont {Faessler}},\ }\href {\doibase 10.1103/PhysRevC.60.055502}
  {\bibfield  {journal} {\bibinfo  {journal} {Phys. Rev. C}\ }\textbf {\bibinfo
  {volume} {60}},\ \bibinfo {pages} {055502} (\bibinfo {year}
  {1999})}\BibitemShut {NoStop}%
\bibitem [{NND()}]{NNDC}%
  \BibitemOpen
  \href {http://www.nndc.bnl.gov/} {\bibinfo  {journal} {National Nuclear Data
  Center (NNDC)}\ ,\ \bibinfo {pages} {http://www.nndc.bnl.gov/}}\BibitemShut
  {NoStop}%
\bibitem [{\citenamefont {Borrajo}\ \emph {et~al.}(2015)\citenamefont
  {Borrajo}, \citenamefont {Rodriguez},\ and\ \citenamefont
  {Egido}}]{Borrajo15}%
  \BibitemOpen
\bibfield  {journal} {  }\bibfield  {author} {\bibinfo {author} {\bibfnamefont
  {M.}~\bibnamefont {Borrajo}}, \bibinfo {author} {\bibfnamefont {T.~R.}\
  \bibnamefont {Rodriguez}}, \ and\ \bibinfo {author} {\bibfnamefont {J.~L.}\
  \bibnamefont {Egido}},\ }\href {\doibase
  http://dx.doi.org/10.1016/j.physletb.2015.05.030} {\bibfield  {journal}
  {\bibinfo  {journal} {Phys. Lett. B}\ }\textbf {\bibinfo {volume} {746}},\
  \bibinfo {pages} {341 } (\bibinfo {year} {2015})}\BibitemShut {NoStop}%
\bibitem [{\citenamefont {Fang}\ \emph {et~al.}(2015)\citenamefont {Fang},
  \citenamefont {Faessler},\ and\ \citenamefont {Simkovic}}]{Fang15}%
  \BibitemOpen
  \bibfield  {author} {\bibinfo {author} {\bibfnamefont {D.-L.}\ \bibnamefont
  {Fang}}, \bibinfo {author} {\bibfnamefont {A.}~\bibnamefont {Faessler}}, \
  and\ \bibinfo {author} {\bibfnamefont {F.}~\bibnamefont {Simkovic}},\ }\href
  {\doibase 10.1103/PhysRevC.92.044301} {\bibfield  {journal} {\bibinfo
  {journal} {Phys. Rev. C}\ }\textbf {\bibinfo {volume} {92}},\ \bibinfo
  {pages} {044301} (\bibinfo {year} {2015})}\BibitemShut {NoStop}%
\bibitem [{\citenamefont {Barea}\ and\ \citenamefont
  {Iachello}(2009)}]{Barea2009}%
  \BibitemOpen
  \bibfield  {author} {\bibinfo {author} {\bibfnamefont {J.}~\bibnamefont
  {Barea}}\ and\ \bibinfo {author} {\bibfnamefont {F.}~\bibnamefont
  {Iachello}},\ }\href {http://dx.doi.org/10.1103/PhysRevC.79.044301}
  {\bibfield  {journal} {\bibinfo  {journal} {Phys. Rev. C}\ }\textbf {\bibinfo
  {volume} {79}},\ \bibinfo {pages} {044301} (\bibinfo {year}
  {2009})}\BibitemShut {NoStop}%
\bibitem [{\citenamefont {Vaquero}\ \emph {et~al.}(2013)\citenamefont
  {Vaquero}, \citenamefont {Rodr{\'\i}guez},\ and\ \citenamefont
  {Egido}}]{Vaquero2013}%
  \BibitemOpen
  \bibfield  {author} {\bibinfo {author} {\bibfnamefont {N.~L.}\ \bibnamefont
  {Vaquero}}, \bibinfo {author} {\bibfnamefont {T.~R.}\ \bibnamefont
  {Rodr{\'\i}guez}}, \ and\ \bibinfo {author} {\bibfnamefont {J.~L.}\
  \bibnamefont {Egido}},\ }\href
  {http://link.aps.org/doi/10.1103/PhysRevLett.111.142501} {\bibfield
  {journal} {\bibinfo  {journal} {Phys. Rev. Lett.}\ }\textbf {\bibinfo
  {volume} {111}},\ \bibinfo {pages} {142501} (\bibinfo {year}
  {2013})}\BibitemShut {NoStop}%
\bibitem [{\citenamefont {Hinohara}\ and\ \citenamefont
  {Engel}(2014)}]{Hinohara14}%
  \BibitemOpen
  \bibfield  {author} {\bibinfo {author} {\bibfnamefont {N.}~\bibnamefont
  {Hinohara}}\ and\ \bibinfo {author} {\bibfnamefont {J.}~\bibnamefont
  {Engel}},\ }\href {\doibase 10.1103/PhysRevC.90.031301} {\bibfield  {journal}
  {\bibinfo  {journal} {Phys. Rev. C}\ }\textbf {\bibinfo {volume} {90}},\
  \bibinfo {pages} {031301} (\bibinfo {year} {2014})}\BibitemShut {NoStop}%
\bibitem [{\citenamefont {Men\'endez}\ \emph {et~al.}(2016)\citenamefont
  {Men\'endez}, \citenamefont {Hinohara}, \citenamefont {Engel}, \citenamefont
  {Mart\'{\i}nez-Pinedo},\ and\ \citenamefont {Rodr\'{\i}guez}}]{Menendez2015}%
  \BibitemOpen
  \bibfield  {author} {\bibinfo {author} {\bibfnamefont {J.}~\bibnamefont
  {Men\'endez}}, \bibinfo {author} {\bibfnamefont {N.}~\bibnamefont
  {Hinohara}}, \bibinfo {author} {\bibfnamefont {J.}~\bibnamefont {Engel}},
  \bibinfo {author} {\bibfnamefont {G.}~\bibnamefont {Mart\'{\i}nez-Pinedo}}, \
  and\ \bibinfo {author} {\bibfnamefont {T.~R.}\ \bibnamefont
  {Rodr\'{\i}guez}},\ }\href {\doibase 10.1103/PhysRevC.93.014305} {\bibfield
  {journal} {\bibinfo  {journal} {Phys. Rev. C}\ }\textbf {\bibinfo {volume}
  {93}},\ \bibinfo {pages} {014305} (\bibinfo {year} {2016})}\BibitemShut
  {NoStop}%
\end{thebibliography}
%

%
%



\end{document}